\documentclass[12pt]{article}
\usepackage{amsfonts,amsmath}
\usepackage{epsfig}
\usepackage{sc3conf}

\makeatletter \@addtoreset{equation}{section} \makeatother

\def\p{\partial}
\def\dt{d\tau}
\def\ds{d\sigma}
\def\ra{\rightarrow}
\def\lra{\longrightarrow}
\def\dox{\dot{x}}
\def\ra{\rightarrow}

\newcommand{\be}{\begin{equation}}
\newcommand{\ee}{\end{equation}}
\newcommand{\ba}{\begin{align}}
\newcommand{\ea}{\end{align}}

\newenvironment{example}[1]{\textbf{Example:} #1}{}

\begin{document}

\vspace{ -3cm}



\begin{center}
\vspace{0.1cm}
{
\Large\bf
Introductory Lectures on String Theory\footnote{These lectures were prepared in collaboration with 
Alexei S. Matveev.}
\vspace{0.3cm}
 }

 \vspace{.2cm} {   A.A.    Tseytlin\footnote{Also at  
 Department of Theoretical Physics, Lebedev Instititute, Moscow, Russia}}

\vspace{.1cm}

{\em 
  Blackett Laboratory, Imperial College,
London SW7 2AZ, U.K.
 }

\end{center}

 \begin{abstract}
We give an elementary introduction to classical and quantum bosonic string theory. 
\end{abstract}

\vskip 2cm

\section{Introduction}

\begin{center}
particles $\longrightarrow$ strings (1-$d$ objects)
\end{center}
\begin{enumerate}
  \item ``effective'' strings: 1-$d$ vortices, solitons, Abrikosov 
   vortex in superconductors, 
   ``cosmic strings''
  in gauge theory; these are  built out of ``matter'', have  finite ``width'', massive
  excitations (including longitudinal ones).

  \item ``fundamental''strings: no internal structure (zero ``thickness''), admit 
  consistent quantum
  mechanical and relativistic description.

\begin{minipage}[h]{0.5\linewidth}
\centering{\epsfig{figure=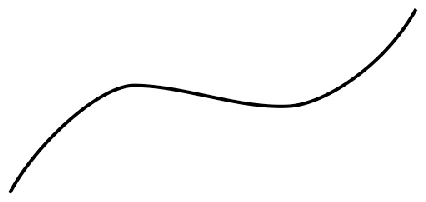,width=4cm,height=3cm}}

Fig.1: Open string
\vspace{0.5cm}
\end{minipage}
\hfill
\begin{minipage}[h]{0.5\linewidth}
\centering{\epsfig{figure=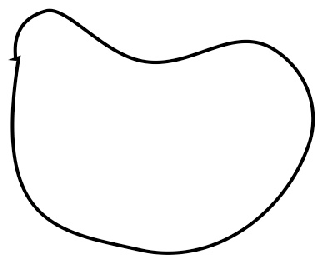,width=4cm,height=3cm}}

Fig.2: Closed string
\vspace{0.5cm}
\end{minipage}
\end{enumerate}

Tension = $\frac{\text{mass}}{\text{length}}=T$ - main parameter of fundamental strings.

\begin{center}
String vibration modes $\longrightarrow$ quantum particles

discrete spectrum of excitations ($\infty$ number)
\end{center}

\be
m^2=-p^2+T  N, \qquad N=0,1,2,\ldots,  \quad p^2=-p_0^2+p_i^2
\ee

\begin{example}{straight open relativistic string rotating about c.o.m.}\par

\vspace{0.3cm}
\begin{minipage}[h]{0.4\linewidth}
\centering{\epsfig{figure=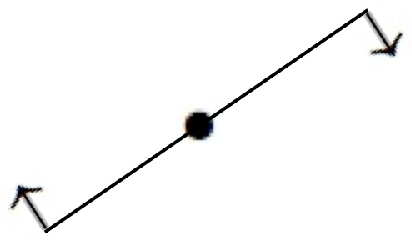,width=4cm,height=3cm}}

Fig.3: Rotating string
\vspace{1cm}
\end{minipage}
\hfill
\begin{minipage}[h]{0.6\linewidth}
length $L$\par
mass $M=T\cdot L$\par
angular momentum $J \sim P \cdot L$\par $P \sim Mc$
(relativistic)\par $J \sim P\cdot L \sim M\cdot L \sim T^{-1} M^2$
\vspace{0.5cm}\par $\boxed{M^2\sim T\cdot J}$ \hspace{0.5cm}
property of relativistic string\par \vspace{0.5cm} Angular moment
$J$ is quantized in QM \vspace{0.5cm}
\vspace{0.2cm}
\end{minipage}
\end{example}

\begin{flushleft}
{\large Why strings?}
\end{flushleft}

\begin{itemize}
  \item consistent quantum theory of gravity and other interactions 
  \item effective description of strongly interacting 
  gauge theories
\end{itemize}

\begin{flushleft}
{\large Why not membranes (or $p$-branes, $p \ge 2$)?}
\end{flushleft}

No consistent QM theory of extended objects with $p>1$ is known

\begin{flushleft}
{\large Quantum theory of gravitation:}
\end{flushleft}

point-like particle (e.g., graviton): interactions are local
$\rightarrow$ UV divergences $\rightarrow$\par non-renormalizable
(need $\infty$ number of counterterms)

\begin{flushleft}
{\large Graviton scattering amplitudes}
\end{flushleft}
\begin{minipage}[h]{0.6\linewidth}
UV divergences $\rightarrow$ no consistent description\par  of
gravity  at short distance

\vspace{0.1cm}
\end{minipage}
\hfill
\begin{minipage}[h]{0.3\linewidth}
\centering{\epsfig{figure=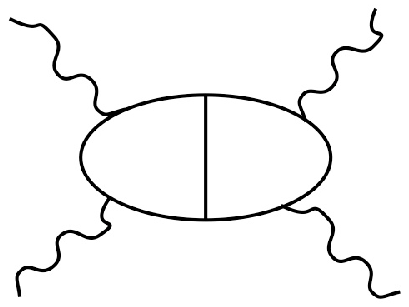,width=4cm,height=3cm}}

Fig.4: Scattering
\vspace{0.1cm}
\end{minipage}

\begin{flushleft}
{\large String interaction are effectively non-local}
\end{flushleft}

\begin{minipage}[h]{0.3\linewidth}
\centering{\epsfig{figure=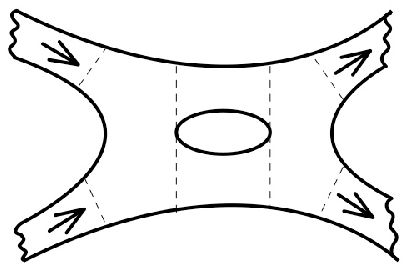,width=4cm,height=3cm}}

Fig.5: 2-$d$ surfaces as ``world-sheets''
\vspace{0.3cm}
\end{minipage}
\hfill
\begin{minipage}[h]{0.6\linewidth}
loop amplitudes are finite $\rightarrow$ consistent in QM 
\vspace{0.5cm}\par graviton  and other ``light''  particles
appear as particular string states 
\vspace{0.3cm}
\end{minipage}

UV finiteness: \quad string scale $\sim$ (tension)$^{-1/2}$   plays the role of
effective cut-off

\begin{flushleft}
{\large Historical  origin -- in the theory of strong interactions
(1968-72)}
\end{flushleft}
\begin{minipage}[h]{0.3\linewidth}
\centering{\epsfig{figure=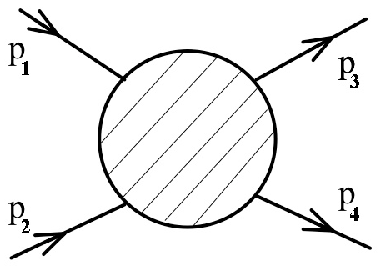,width=4cm,height=3cm}}

Fig.6: Scattering of hadrons
\vspace{0.3cm}
\end{minipage}
\hfill
\begin{minipage}[h]{0.6\linewidth}
$s=-(p_1+p_2)^2$ \vspace{0.3cm}\par $t=-(p_3+p_4)^2$
\vspace{0.3cm}\par $A(s,t)$ = hadron amplitude \vspace{0.3cm}\par duality
observed: $A(s,t)=A(t,s)$
\vspace{0.3cm}
\end{minipage}
\begin{minipage}[h]{0.5\linewidth}
Unusual for field theories\vspace{0.3cm}\par%
$\mathcal{L}=(\partial\phi)^2+m^2\phi^2+\lambda\phi^3$
\vspace{0.1cm}
\end{minipage}
\hfill
\begin{minipage}[h]{0.4\linewidth}
\centering{\epsfig{figure=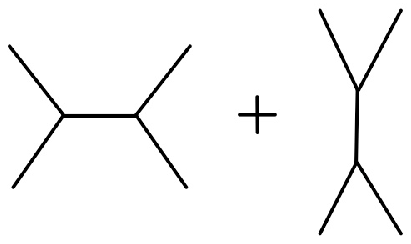,width=4cm,height=3cm}}

Fig.7: $s,t$ channels are not the same
\vspace{0.1cm}
\end{minipage}
\begin{minipage}[h]{0.3\linewidth}
\centering{\epsfig{figure=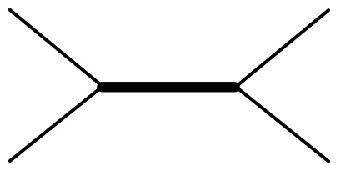,width=4cm,height=3cm}}

Fig.8: Resonances
\vspace{0.1cm}
\end{minipage}
\hfill
\begin{minipage}[h]{0.6\linewidth}
But possible in a theory with $\infty$ number of ``resonances''
(intermediate states)\vspace{0.3cm}\par $m^2 \sim n=1,2,\ldots$
\vspace{0.1cm}
\end{minipage}

\vspace{0.2cm}
\begin{minipage}[h]{0.3\linewidth}
\centering{\epsfig{figure=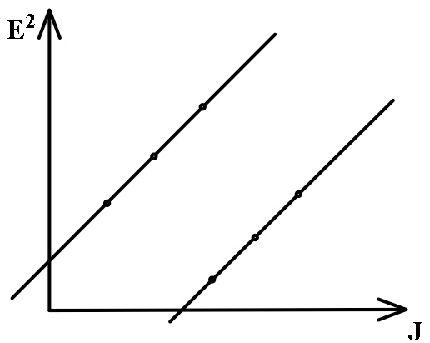,width=4cm,height=3cm}}

Fig.9: Regge trajectories
\vspace{0.1cm}
\end{minipage}
\hfill
\begin{minipage}[h]{0.6\linewidth}
Linear relations between energy-squared  $E^2$ of resonances and their
spins $J$  characterstic to string theory spectrum 
were indeed observed in experiments. 
\vspace{0.8cm}
\end{minipage}

\section{Classical String Theory}

\begin{flushleft}
{\large Point-like particle}
\end{flushleft}

\begin{minipage}[h]{0.3\linewidth}
\centering{\epsfig{figure=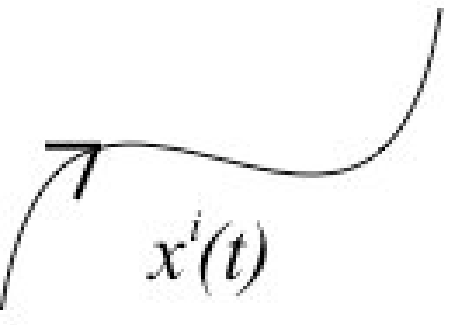,width=4cm,height=3cm}}

Fig.10: Non-relativistic massive particle
\vspace{0.5cm}
\end{minipage}
\hfill
\begin{minipage}[h]{0.6\linewidth}
\be
\mathcal{L}=\sum_{n=1}^{D-1}\frac{m\dot{x}_n^2}{2}\ 
\ee
\vspace{0.5cm}
\end{minipage}
\noindent
Action for a non-relativistic massive particle (assume no higher derivatives)\footnote{In what
follows we will omit summation  sign so that $\sum_n a_n^2 \equiv a_n^2
\equiv a^2$.}:
\begin{equation}
    S=-mc^2\int dt \sqrt{1-\frac{\dot{x}_n^2}{c^2}} \simeq
    \int dt \left( -mc^2 + \frac{m\dot{x}^2_n}{2} + \ldots \right),
    \qquad \dot{x} \ll c
\end{equation}
Momentum:
\begin{equation}
    p_n =\frac{\partial \mathcal{L}}{\partial x^n} =
     \frac{m\dot{x}_n}{\sqrt{1-\frac{\dot{x}^2}{c^2}}}
\end{equation}
Invariance of the action: Lorentz transformations

Manifestly relativistic-invariant form (in what follows $c=1$)
\begin{equation}
    S= -m\int d\tau \sqrt{-\dot{x}^\mu\dot{x}^\nu\eta_{\mu\nu}}, \qquad
    \mu=(0,n)=(0,1,\ldots,D-1)
\end{equation}
$\eta_{\mu\nu}=$diag(-1,1,...,1), 
$D$ is the dimension of Minkowski space-time $\{x^\mu\}$,
$\tau$ is a world-line parameter.

Equivalently,
\begin{equation}
    S= -m\int ds,
\end{equation}
where
\begin{equation} \notag
    ds^2=\eta_{\mu\nu}dx^\mu dx^\nu, \quad dx^\mu =
    \frac{dx^\mu}{d\tau}d\tau
\end{equation}
\begin{equation}
    S= -m\int d\tau \sqrt{(\dox^0)^2-(\dox^n)^2}
\end{equation}
Symmetries:
\begin{enumerate}
  \item Space-time: Poincare Group
  \be
  x'^{\mu} = \Lambda^\mu{}_\nu x^\nu + a^\mu, \quad \Lambda \in SO(1,D-1)
  \ee
\be
\eta_{\mu\nu} \Lambda^\mu{}_{\mu'} \Lambda^\nu{}_{\nu'} =
\eta_{\mu'\nu'} \quad \Rightarrow \quad (x,y)=\eta_{\mu\nu}x^\mu
y^\nu = (x',y')
\ee
$$
ds^2=\eta_{\mu\nu}dx^\mu dx^\nu = \text{invariant}
$$
  \item World-line: reparametrization invariance
  $$
\tau'=f(\tau), \qquad x'^\mu(\tau') = x^\mu(\tau)
  $$
$$
\tau'=\tau +\xi(\tau), \qquad  x'=x+\delta x \quad \Rightarrow \quad \delta
x^\mu = -\xi \dox^\mu
$$
\end{enumerate}

``Proper-time'' or ``static''  gauge (special coordinate system in 1-$d$)

$x^0(\tau)=\tau \equiv t$ --- one function is fixed

$x'^0(f(\tau))=x^0(\tau)=\tau \longrightarrow$ fixes  $f(\tau)$

(cf. $A_0=0$ gauge in Maxwell theory)

$\sqrt{(\dot{x}^0)^2-(\dot{x}^n)^2} \longrightarrow
\sqrt{1-(\dot{x}^n)^2} = $ usual relativistic action

Action in static gauge
\be
S=-m \int dt \sqrt{1-\dot{x}_n^2}
\ee
Equivalently, changing $\tau \rightarrow x^0(\tau)$
\be
S=-m \int dt
\sqrt{(\frac{dx^0}{d\tau})^2-(\frac{dx^n}{d\tau})^2} =
-m \int dx^0 \sqrt{1-(\frac{dx^n}{dx^0})^2}
\ee

\subsection{String action}

\begin{flushleft}
{\large Non-relativistic string}
\end{flushleft}
\begin{minipage}[h]{0.4\linewidth}
\centering{\epsfig{figure=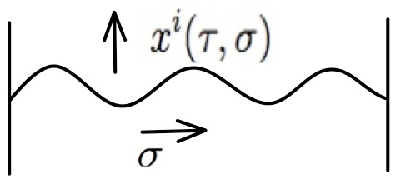,width=4cm,height=3cm}}

Fig.11: Non-relativistic string
\vspace{0.3cm}
\end{minipage}
\hfill
\begin{minipage}[h]{0.5\linewidth}
small $\perp$  oscillations ($i=1,..., D-2$)
$$
\mathcal{L} \approx  \frac12 T(\dot{x}_i^2+x'^2_i)
$$
$T$ -- $\text{tension}=\frac{\text{mass}}{\text{length}}$
\vspace{0.3cm}
\end{minipage}

\noindent
world surface coordinates $x^i(\tau,\sigma)$ (here $\tau \equiv x^0=t$)

\noindent
``transverse'' string coordinates $\dot{x}_i=\frac{\p x_i}{\p t},
\  x'_i=\frac{\p x_i}{\p \sigma}$
\be
S \approx  \frac12 T  \int d\tau \int d\sigma \ (\dot{x}_i^2+x'^2_i)
\ee
Center of mass: $x_i(\tau,\sigma)=\overline{x}_i (\tau)+\ldots$
\be
S \rightarrow   \frac12 T   L \int d\tau\ \dot{\bar x}_i^2+\ldots =
\frac{m}{2}\int d\tau\ \dot{\bar x}_i^2+\ldots
\ee

\begin{flushleft}
{\large Relativistic  string}
\end{flushleft}
Basic principles:
\begin{enumerate}
  \item Relativistic invariance (Poincar\'e   group)
  \item No internal structure, i.e. 
  no longitudinal oscillations $\rightarrow$ 
  2-$d$ reparametrization invariance
\end{enumerate}
Relativistic generalization: $x^i \rightarrow x^\mu$

The Nambu-Goto  action\footnote{Y. Nambu, Lectures at the Copenhagen Symposium, 1970;
 T.~Goto,
  ``Relativistic quantum mechanics of one-dimensional mechanical continuum and
  subsidiary condition of dual resonance model,''
  Prog.\ Theor.\ Phys.\  {\bf 46}, 1560 (1971).
    }
 is  (we assume no higher than first derivatives are present in the action)
\be
S= -T \int d\tau \int d\sigma \sqrt{(\dox  x')^2-\dox^2 x'^2}
\ee
Motion in space-time: $\{x^\mu(\tau,\sigma)\}$, \  $(\tau,\sigma)$ --
 world-surface coordinates
\be
\dox  x'\equiv \dox^\mu x'^\nu\eta_{\mu\nu}, \qquad \dox^2\equiv 
\dox^\mu x'^\nu\eta_{\mu\nu}
\ee
Symmetries:
\begin{enumerate}
  \item space-time (global) --- Poincar\'e: \  $x'^\mu=\Lambda^\mu{}_{\nu}x^\nu+a^\mu$
  \item 2-$d$ world-volume (local) --- reparametrizations of world-surface
\end{enumerate}
$$
\tau'=f(\tau,\sigma), \qquad \sigma'=g(\tau,\sigma)
$$
with 
$x^\mu(\tau,\sigma)$ transforming  as   scalars in 2-$d$:
\be
x'^\mu(\tau',\sigma')=x^\mu(\tau,\sigma)
\ee
$$
\xi^a=(\tau,\sigma), \qquad \xi'^a=\xi^a+\zeta^a(\xi)
$$
$$
x'^\mu(\xi+\zeta)=x^\mu(\xi) \longrightarrow \delta x^\mu =-\zeta^a \p_a x^\mu,
\qquad \p_a=(\p_0,\p_1)=(\p_\tau,\p_\sigma)
$$

Meaning of the action:\ area of world surface
\begin{center}
\epsfig{figure=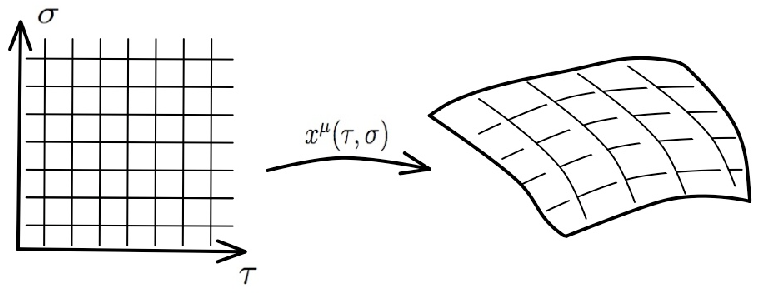,width=8cm,height=3.5cm}

Fig.12: Embedding of ($\tau$,$\sigma$)-plane into space-time
\end{center}

\vspace{0.1cm}
\noindent Induced metric on world surface  $(a,b=0,1)$
\be
ds^2=\eta_{\mu\nu}dx^\mu dx^\nu = h_{ab}(\xi)d\xi^a d\xi^b
\ee
\be
h_{ab}=\p_a x^\mu \p_b x^\nu \eta_{\mu\nu}
\ee
\be
h_{ab}=\left(
  \begin{array}{cc}
    \dox^2 & \dox x' \\
    \dox x' & x'^2 \\
  \end{array}
\right), \qquad \det h_{ab} = \dox^2 x'^2 - (\dox x')^2
\ee
Thus Nambu-Goto action reads
\be
S=-T \int d^2\xi\ \sqrt{-h}, \qquad h = \det h_{ab}
\ee
Euclidean signature ($\eta_{\mu\nu} \rightarrow \delta_{\mu\nu}$)
\be
\tau \ra i \tau_E, \qquad ds^2=-\dt^2+\ds^2 \ra \dt^2_E+\ds^2
\ee
\be
S_E = -iS_M = T \int \dt_E \int \ds \sqrt{\dox^2 x'^2-(\dox x')^2}
\ee
Element of  area:
$$
d(\text{Area})=|\dox||x'|\sin\alpha \,\ds \dt_E
$$
$$
\cos\alpha = \frac{\dox x'}{|\dox||x'|}\ , \qquad \sin\alpha=\sqrt{1-\cos^2\alpha}
$$
\begin{center}
\epsfig{figure=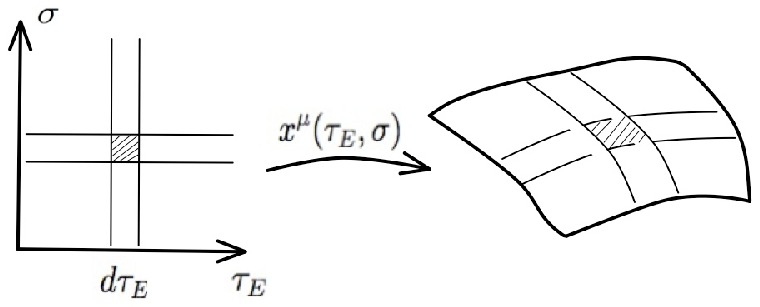,width=9cm,height=4cm}

Fig.13: Element of area
\end{center}
\vspace{0.3cm}
\be
(\text{Area})_E = \int \dt_E \ds \sqrt{|\dox|^2|x'|^2}\sqrt{1-\frac{(\dox x')^2}{|\dox|^2|x'|^2}}
=\int \dt_E \ds \sqrt{\dox^2 x'^2-(\dox x')^2}
\ee
Static gauge:
$$
x^0=\tau,\quad x^{D-1}=\sigma, \quad x^i(\tau,\sigma)=\text{transverse coordinates}, \quad i=1,\ldots, D-2
$$
\be
h_{ab}=\p_a x^\mu \p_b x^\nu \eta_{\mu\nu}=\eta_{ab}+\p_a x^i \p_b x^i
\ee
Expand action assuming $|\p_a x^i| \ll 1$ (small oscillations):
\be
S=-T\int d^2\xi \sqrt{-\det(\eta_{ab}+\p_a x^i \p_b x^i)}
\ee
$$
\det \eta_{ab}=-1
$$
$$
\det(\eta_{ab}+\p_a x^i \p_b x^i)=\det \eta_{ab} \det(\delta_d^c+\p_d x^i \p_b x^i \eta^{bc})
$$
$$
\simeq -(1+\p_a x^i \p_b x^i \eta^{ab}+\ldots)
$$
$$
S=-T\int d^2\xi \big( 1+\frac12 \p^a x^i \p_a x^i\big)+ \mathcal{O}((\p x)^4)
$$
$$
\simeq -m\int\dt+\frac12 T\int \dt \int_{0}^{L}\ds \big[  (\dox^i)^2-(x'^i)^2\big]+ \mathcal{O}((\p x)^4)
$$
Equation for small oscillations:\newline
$\ddot{x}_i-x''_i=0 \lra$ wave equation in 1-$d$:  transverse waves on the string

\subsection{Relation to particle action}

Nambu-Goto action describes 
string as collection of particles moving in  direction  transverse to the 
string.
\newline
Indeed, use static gauge for $\tau$ only  ($x^0=\tau$) and start with: 
\be
S=-T\int \dt \int dl \sqrt{1-V^2_{\bot}},
\ee
where   
\be
dl \equiv \ds \sqrt{|x'|^2}, \qquad  V^n_{\bot}x'^n=0
\ee
and  $x\equiv (x^n(\tau,\sigma)), \  n=1,\ldots,D-1$.
Solution of $V^n_{\bot} x'_n=0: $
\be
V^m_\bot = P^{mn}\dox^n \ , \ \ \ \ \ \ \ \ 
P^{mn}=\delta^{mn}-\frac{x'^m x'^n}{|x'|^2}, \qquad P^{mn}x'^n=0
\ee
\be
V^2_\bot \equiv V^n_\bot V^n_\bot = P^{nm}P^{nk}\dox^m \dox^k = \dox^2-\frac{(\dox x')^2}{x'^2}
\ee
\be
S=-T\int\dt\int dl\sqrt{1-V^2_\bot}= -T\int\dt\int\ds \sqrt{x'^2+(\dox x')^2-\dox^2 x'^2}
\ee
\newline
This action (in $x^0=\tau$ gauge)
is invariant under $\sigma \ra f(\sigma)$.
\newline
This coincides   with the Nambu-Goto action in the  ``incomplete'' static gauge  $x^0=\tau$:
$$
-\det h_{ab} = -(\dox_\mu \dox^\mu)(x'^\nu x'_\nu) +(\dox^\mu x'_\mu)^2
= (x'_n)^2-(x'_n)^2(\dox_m)^2 +(\dox_n x'_n)^2
$$
The Nambu-Goto action
thus 
 describes a collection of particles ``coupled'' by 
 the constraint that they should move
transversely to the profile of the string (i.e. there should be 
no longitudinal motions).

\subsection{$p$-brane action}
\begin{itemize}
  \item $p=0$: particle
  \item $p=1$: string
  \item $p=2$: membrane (2-brane), etc.
\end{itemize}
$p$-brane --- $(p+1)$-dim world surface $\Sigma^{p+1}$ embedded in Minkowski space $\mathbb{M}^D$
\newline
Embedding: $x^\mu(\xi^a),\quad\mu=1,\ldots,D-1, \quad a=0,\ldots,p$

Principles:
\begin{enumerate}
  \item  Poincare invariance  (global)
  \item  world-volume reparametrization invariance  (local)
  \item  no higher derivative (``acceleration'') terms
\end{enumerate}
$$
\mathcal{L} \simeq -\frac12 T_p \p_a x^i \p_b x^i \eta^{ab}+\ldots
$$
where $\eta^{ab}=(-1,+1,\ldots,+1)$
\newline
Static gauge:
\be
x^0=\xi^0,\ldots,\  x^p=\xi^p; \qquad x^i(\xi)=x^i(\xi^0,\ldots,\xi^p)
\ee
$x^i(\xi)$ --- dynamical (transverse) coordinates

Action that satisfies reparametrization invariance  condition 
 has geometric interpretation:
volume of induced metric on $\Sigma^{p+1}$
\be
ds^2=\eta_{\mu\nu}dx^\mu dx^\nu = h_{ab}(x(\xi))d\xi^a d\xi^b,
\qquad h_{ab} = \p_a x^\mu \p_b x^\nu \eta_{\mu\nu}
\ee
\begin{equation}\label{Sp(h)-action}
   S_p = -T_p \int d^{p+1}\xi \sqrt{-h}, \qquad h=\det h_{ab}
\end{equation}
Poincar\'e and reparametrization invariance are obvious.
\newline
Static gauge: use first $(p+1)$ coordinates 
$x^0,\ldots, x^p$ to parametrise the surface
$$
x^a=\xi^a, \qquad x^i \equiv (\tilde{x}^1,\ldots,\tilde{x}^{D-1-p})\equiv (x^{p+1},\ldots, x^{D-1})
$$
$$
h_{ab}=\eta_{ab}+\p_a x^i \p_b x^i
$$
$$
\det h_{ab} = -(1+\eta^{ab} \p_a x^i \p_b x^i + \ldots)
$$
\be \label{Spx}
S_p = -T_p \int d^{p+1}\xi \sqrt{1+\p^a x^i \p_a x^i +\ldots}
= -T_p \int d^{p+1}\xi (1+\frac12 \p^a x^i \p_a x^i+\ldots)
\ee
For $p=0,1$ one can  find a gauge in which the  equations 
of motion become linear; but this is not possible  for $p \ge 2$, i.e.  
the  equations  are always nonlinear.
\newline
Variational principle: $\quad x^\mu \ra x^\mu + \delta x^\mu$
\be
\delta S_p= -T_p \int d^{p+1}\xi\ \delta \sqrt{-\det h_{ab}(x(\xi))}
\ee
\be
\delta h_{ab} =\eta_{\mu\nu} \p_a x^\mu \p_b (\delta x^\nu) + (a \leftrightarrow b), \qquad
\delta \sqrt{-h} = \frac12 \sqrt{-h}h^{ab} \delta h_{ab}
\ee
Used  that  $\det A = e^{Tr \ln A}$.  Then 
\be
\delta S_p =T_p \int d^{p+1}\xi\ \delta x^\mu \p_b(\sqrt{-h}h^{ab}\p_a x_\mu)
-T_p \int d^{p+1}\xi\ \p_b(\sqrt{-h}h^{ab}\p_a x_\mu \delta x^\mu)
\ee
Assume boundary conditions that the boundary term here vanishes; e.g., for $p=1$
\ ($0 \le \sigma \le L$):

\begin{itemize}
  \item  closed string: \  $x^\mu(\sigma)=x^\mu(\sigma+L)$, with the 
  boundary condition
  $\delta x^\mu(\tau_{in})=\delta x^\mu(\tau_{fin})=0$
  \item open string: \   free ends -- Neumann boundary  condition -- $\p_\sigma x^\mu=0$
  at $\sigma=0, L$
\end{itemize}
\begin{flushleft}
{\large Equation   of motion:}
\end{flushleft}
\be\label{tak}
\p_b (\sqrt{-h}h^{ab}\p_a x^\mu)=0
\ee
Highly non-linear;  simplifies  in a  special gauge
 to a linear one  for $p=0,1$ only.

\subsection{Action with  auxiliary metric on the world surface
}

In addition to $x^\mu(\xi)$, let us 
 introduce new auxiliary  field $g_{ab}(\xi)$ --- metric tensor on $\Sigma^{p+1}$.
\newline
Classically equivalent action for $p$-brane is then:
\be \label{Ipxg}
I_p(x,g)=-\frac12 T_p \int d^{p+1}\xi \sqrt{-g} (g^{ab}\p_a x^\mu \p_b x_\mu - c)\ , \ \ \
\ \ \   \ c=p-1 
\ee
Action is now quadratic in $x^\mu$ but it contains independent $g_{ab}(\xi)$
 field.
 We shall call  this ``auxiliary metric action''.\footnote{In the  string ($p=1$, $c=0$) case  this action 
 originally  appeared in the papers 
 [S.~Deser and B.~Zumino,
  ``A complete action for the spinning string,''
  Phys.\ Lett.\  B {\bf 65}, 369 (1976)]
  and 
[L.~Brink, P.~Di Vecchia and P.~S.~Howe,
  ``A locally supersymmetric and reparametrization invariant action for the
  spinning string,''
  Phys.\ Lett.\  B {\bf 65}, 471 (1976)]
which generalized a similar construction  for the 
(super) particle case ($p=0$, $c=-1$)  in 
 [L.~Brink, S.~Deser, B.~Zumino, P.~Di Vecchia and P.~S.~Howe,
  ``Local supersymmetry for spinning particles,''
  Phys.\ Lett.\  B {\bf 64}, 435 (1976)].
An equivalent  action  with independent  Lagrange multiplier fields 
that led to the correct constraints (but did not have an immediate geoemtrical interpretation)
 appeared earlier in 
 [P.A.~Collins and R.W.~Tucker,
  ``An action principle and canonical formalism for the Neveu-Schwarz-Ramond
  string,''
  Phys.\ Lett.\  B {\bf 64}, 207 (1976)]
(see footnote  in the    Brink et al  paper and the discussion at the end of  the Deser and Zumino paper).
 This ``auxiliary metric'' form of the classical string  action
  was widely used     after  Polyakov   
 have chosen  it as a starting point for path integral quantization of the  string  [A.~M.~Polyakov,
  ``Quantum geometry of bosonic strings,''
  Phys.\ Lett.\  B {\bf 103}, 207 (1981)]
 }
Eliminating $g_{ab}$ through its equations of motion yields the same equations
for $x^\mu$ as following from \eqref{Sp(h)-action}.

{Interpretation}: action for scalar fields $\{x^\mu\}$ in $d=p+1$ dimensional space $\Sigma^d$
with metric $g_{ab}(\xi)$.

Variation of \eqref{Ipxg} with respect to $x^\mu$ yields
\be \label{var-x}
\p_a(\sqrt{-g}g^{ab}\p_b x^\mu)=0,
\ee
which is of the same form as for
a set of   scalar fields
 $x^\mu$  of zero mass  in curved space with metric $g_{ab}$.

Recall that variations of the inverse of the metric tensor and 
of the determinant of the metric are 
\be \label{var-g}
\delta g^{ab} = -g^{ac}g^{bd}\delta g_{cd}, \qquad \delta \sqrt{-g} =\frac12 \sqrt{-g}g^{ab}\delta g_{ab} .
\ee
With the help of \eqref{var-g} the variation of \eqref{Ipxg} with respect to $g_{ab}$ reads
\be
\int d^{p+1}\xi \sqrt{-g}\  \delta g_{ab} \left( \frac12 g^{ab} (g^{cd}\p_c x^\mu \p_d x_\mu - c)-
g^{ac}g^{bd}\p_c x^\mu \p_d x_\mu \right)=0.
\ee
Using the notation 
 $h_{ab}=\p_a x^\mu \p_b x_\mu$ and lowering the indices with $g_{ab}$ we obtain
\be
\frac12 g_{ab} (g^{cd}\p_c x^\mu \p_d x_\mu - c) =h_{ab} \ , 
\ee
i.e. 
\be 
g_{ab} = \lambda h_{ab} \ , \ \ \ \ \ \ \ \ \ \ 
\lambda^{-1} =\frac12 (g^{ab}h_{ab} - c). 
\ee
On the other hand,
\be
g^{ab} h_{ab} =  \lambda^{-1} g^{ab} g_{ab} = \lambda^{-1} (p+1)\ ,
\ee
i.e.  
\be
c=\lambda^{-1}(p-1).
\ee
Thus, if $p \ne 1$ and $c=p-1$ we get $\lambda=1$, or $g_{ab}=h_{ab}$.
 This means that on the equations of motion the  independent metric $g_{ab}$
 coincides with the induced metric (for $p=1$  it is proportional to it up to an arbitrary 
  factor).

Then equation for $x^\mu$ in \eqref{tak}  becomes
\be \label{var-xx}
\p_a(\sqrt{-h}h^{ab}\p_b x^\mu)=0,
\ee
which is the same as following from the original $S_p[x]$ action \eqref{Sp(h)-action}.

Thus, $I_p[x,g]$ and $S_p[x]$ give the same equations of motion and also are equal
 if we eliminate $g_{ab}$ using the  equation of motion $g_{ab}=h_{ab}$
\begin{align*}
\left.I_p[x,g] \right|_{_{g_{ab}=h_{ab}}} &= -\frac12 T_p \int d^{p+1}\xi \sqrt{-h} (h^{ab}\p_a x^\mu \p_b x_\mu -c) \notag\\
&= -\frac12 T_p \int d^{p+1}\xi \sqrt{-h} (h^{ab}h_{ab}-p+1) \notag\\
& =-\frac12 T_p \int d^{p+1}\xi \sqrt{-h}  = S_p[x].
\end{align*}

\subsection{Special cases}


The tensor $g_{ab}$ is symmetric, so 
it has $(p+1)\times (p+1)$ components which are 
functions of $(p+1)$ arguments  $\xi^a$. Reparametrization
 invariance (gauge freedom) is described with $(p+1)$ 
 functions $\xi'^a=f^a(\xi)$. Thus  the
  number of non-trivial components of $g_{ab}$ equals 
  to $\frac{p(p+1)}{2}$.

\subsubsection{$p=0$:\  particle}

In this case $a,b=0$ and we have only one metric component
 $g_{ab}=g_{00}=-e^2$. Using the following notation $\xi^a=\tau,
 \ T_0=m,\ e=e(\tau)$ we can rewrite \eqref{Ipxg} in the form
\be
I_0[e,x]=-\frac12T_0\int d\tau\ e\ (-e^{-2}\dot{x}^\mu\dot{x}_\mu +1)=\frac12m\int d\tau\
 (e^{-1}\dot{x}^2-e)
\ee
Variation with respect to $e$ gives 
$$
e^{-2} \dot{x}^2+1=0\ ,  \qquad \text{i.e.} \qquad e=\sqrt{-\dot{x}^2}\ .
$$
Then
\be
\left.I_0\right|_{e=\sqrt{-\dot{x}^2}}=-m\int d\tau \sqrt{-\dot{x}^2}=S_0[x].
\ee
Rescaling $e$ by $m$, i.e.  $e(\tau)=m \varepsilon(\tau)$,  leads to 
an action that admits a regular 
 massless limit $m \rightarrow 0$:
\be \label{m-msl}
I_0=\frac12 \int d\tau (\varepsilon^{-1}\dot{x}^2-m^2 \varepsilon) \ .
\ee
The limit $m=0$ gives an   action for the massless relativistic particle
\be
I^{(0)}_0[\varepsilon,x]=\frac12\int d\tau\ \varepsilon^{-1}\dot{x}^2.
\ee
Variation of \eqref{m-msl} with respect to $x^\mu$ and $\varepsilon$ gives
the  equations of motion
\begin{align}
&\frac{d}{d\tau}\left( \varepsilon^{-1} \dot{x}^\mu \right) = 0 \\
& \dot{x}^\mu\dot{x}_\mu = -m^2 \varepsilon^2.
\end{align}
Here the number of gauge parameters equals to 1
and is 
 the same as number of components of $g_{ab}$.
  Thus $g_{00}=- m^2 \varepsilon^2$  can be completely gauged away.
Choosing the  special reparametrization gauge 
$$\varepsilon(\tau)=1$$
 gives 
\be
\ddot{x}^\mu=0 \qquad \rightarrow \qquad x^\mu=x_0^\mu +p^\mu\tau,
\ee
and 
\be
\dot{x}^\mu \dot{x}_\mu=-m^2 \qquad \rightarrow \qquad p^\mu p_\mu =-m^2.
\ee
These are the usual relativistic particle relations.

\subsubsection{$p=1$:\  string}

This  is also  a special case because $(p-1)\lambda^{-1}=c=0$ 
is satisfied identically for any $\lambda$.
Thus, solution is $g_{ab}=\lambda h_{ab}$, where $\lambda=\lambda(\xi)$ 
is an arbitrary function of $\xi$.

The equations for $x^\mu$ are still the same  for the actions  $S_p$ and $I_p$ since 
\be
\sqrt{-g}g^{ab} =\sqrt{-h}h^{ab}.
\ee
This is due to an extra symmetry
that  appears in this case. Indeed, in 2 dimensions ($d=p+1=2$) the action ($p=1,c=0, \ T=
T_1$)
\be
I_1[x,g] = -\frac12 T\int d^2 \xi \sqrt{-g} g^{ab} \p_a x^\mu \p_b x_\mu
\ee
is invariant under the Weyl (or ``conformal'') transformations
\be
g'_{ab} = f(\xi) g_{ab}.
\ee
This symmetry is present only for massless scalars in 
 $d=2=p+1$. For any $p$
\be
\sqrt{-g'}g'^{ab} =f^{\frac{d-2}{2}}\sqrt{-g} g^{ab}.
\ee
This is another 
local (gauge) symmetry in addition to the 
 reparametrization invariance.
  $g_{ab}$ has 3 components -- same as the number
    of gauge parameters of the Weyl (one) and the reparametrization invariance (two) 
   transformations. This 
    allows one  to gauge away $g_{ab}$ completely,  i.e. to set 
    $g_{ab} \sim \eta_{ab}$, 
    and then obtain a  free action  for $x^\mu$.

For general $p$ the  equation for $g_{ab}$ is  the vanishing of the total energy-momentum tensor
\be
T_{ab} = -\frac{2}{\sqrt{-g}} \frac{\delta I_p}{\delta g^{ab}}= T_p(\p_a x^\mu\p_b
 x_\mu -\frac12 g_{ab}g^{cd}\p_c x^\mu\p_d x_\mu)-\frac{c}{2}T_pg_{ab}.
\ee
For $c=0$ and $g_{ab}=\lambda h_{ab}$  we have $T_{ab}(x)=
T ( \p_a x^\mu\p_b
 x_\mu -\frac12 h_{ab}h^{cd}\p_c x^\mu\p_d x_\mu) =  0$.

\

To summarize, in the $p=1$  case we  have 
the following local gauge transformations
\begin{enumerate}
\item Reparametrization invariance:

\hspace{3.00cm} $\xi'^a=f^a(\xi) \quad \rightarrow \quad x'^\mu(\xi')=x^\mu(\xi),\, \ \ \ 
g'_{ab}(\xi')=
\frac{\p\xi^c}{\p\xi'^a}\frac{\p\xi^d}{\p\xi'^b}g_{cd}(\xi)$ 

\item Weyl invariance: 

\qquad $x'^\mu(\xi)=x^\mu(\xi),\,\ \  \ \ g'_{ab}(\xi)=\sigma(\xi)g_{ab}$
\end{enumerate}
Total number of gauge functions equals to the 
number of components of $g_{ab}$  so that  it can be gauged away completely.

Namely,  one 
can choose the  ``conformal'' (or ``orthogonal'') gauge $g_{ab}= 
\eta_{ab}$.
It is sufficient even to choose the reparametrization gauge  only 
on the 
 Weyl-invariant combination: $\sqrt{-g}g^{ab}=\eta^{ab}$.

Then the equation for $x^\mu$ becomes linear.
Indeed, 
consider the  reparametrization 
 gauge $g_{ab}=f(\xi) \eta_{ab}$, where $f$ is arbitrary.
Then, $g_{00}=-g_{11}$ and $g_{01}=0$  and  from the equation of motion 
\be
\p_a(\sqrt{-g}g^{ab}\p_bx^\mu)=0
\ee
we have
\be
\p^a\p_a x^\mu=0 \ , \qquad \text{i.e.} \qquad \ddot{x}^\mu-x''^\mu=0.
\ee
Also we know that $h_{ab} \sim g_{ab} \sim \eta_{ab}$, where $h_{ab}=\p_a x^\mu \p_b x_\mu$.
Then, $h_{00}=-h_{11}$ and $h_{01}=0$. In other words, we have in addition 
the following
 differential constraints (containing only 1st time derivatives and generalizing 
          $\dot{x}^2_\mu =0$ for a massless particle)
\be
\dot{x}^2_\mu +x'^2_\mu=0 \ , \qquad  \qquad \dot{x^\mu}x'_\mu=0.
\ee

\subsection{Meaning of the constraints and first-order actions
}

\subsubsection{Particle case}

Let us introduce an  independent momentum function 
$p^\mu(\tau)$. Then  the 
action for $\{ x^\mu(\tau), p^\mu(\tau), \varepsilon(\tau)\}$ 
which is 
equivalent to \eqref{m-msl}  is 
\be \label{hatio}
\hat{I}_0(x,p,\varepsilon) = \int d\tau \left[ \dot{x}^\mu p_\mu -\frac12 \varepsilon (p^2+m^2)
   \right] \ ,
\ee
where $\varepsilon$ is a Lagrange multiplier imposing the  constraint $p^2+m^2=0$.

The variation of \eqref{hatio} gives 
\begin{itemize}
\item $\delta x^\mu:$ \qquad $\dot p_\mu=0$
\item $\delta p^\mu:$ \qquad $\dot x^\mu=\varepsilon p^\mu$
\item $\delta \varepsilon:$ \qquad $p^2+m^2=0$
\end{itemize}
This action is invariant under the 
reparametrization with $d\tau \varepsilon(\tau)=d\tau' \varepsilon'(\tau')$. Fixing 
the  gauge as $\varepsilon=1$ we get the usual equations
\be
p^\mu=\dot x^\mu\ ,\ \ \ \  \qquad \ddot x=0.
\ee
Eliminating  $p^\mu$ from the action \eqref{hatio} yields \eqref{m-msl}, i.e. 
\be
\left.\hat{I}_0(x,p,\varepsilon) \right|_{p=\varepsilon^{-1}\dot x} = \frac12 \int d\tau (\varepsilon^{-1}\dot{x}^2-\varepsilon m^2).
\ee
Thus $g_{00}=-m^2\varepsilon^2$ plays the  role of a  Lagrange multiplier.

\subsubsection{String case}

Similarly, let us introduce the independent momentum field
 $p^\mu(\tau,\sigma)$  and consider the 
 alternative action 1-st order action 
 for $x^\mu, p_\mu$ with the conformal-gauge 
  constraints added with the 
  Lagrange multipliers $\varepsilon(\tau,\sigma)$ and 
  $\mu(\tau,\sigma)$\be \label{I1xp}
\hat I_1(x,p,\varepsilon,\mu) = T\int d\tau d\sigma \left[\dot x^\nu p_\nu -\ \frac12
\varepsilon\ 
(p^2+x'^2)-\ \mu \ p_\nu x'^\nu\right]
\ee
The variation of \eqref{I1xp} gives 
\begin{itemize}
\item $\delta x^\nu:$ \qquad $\dot p_\nu - (\varepsilon x'_\nu)'=0$
\item $\delta p^\nu:$ \qquad $p_\nu=\varepsilon^{-1} (\dot x_\nu - \mu x'_\nu)$
\item $\delta \varepsilon:$ \qquad $p^2+x'^2=0$
\item $\delta \mu:$ \qquad $p^\nu  x'_\nu =0$
\end{itemize}
This action is invariant under the 2-parameter 
 reparametrization symmetry 
 $\xi'_a=f_a(\xi),\ \xi=\xi(\tau,\sigma)$. So,  one can fix 
 the two gauges $\varepsilon=1,\  \mu=0$. Then we get 
\be\label{orth}
p_\mu=\dot x_\mu,\  \qquad \ddot x_\mu - x''_\mu=0, 
\qquad \dot x^2+ x'^2 = 0, \qquad \dot x x'=0,
\ee
i.e. the same set of equations as  in the orthogonal gauge.

As in the  particle case, the $g_{ab}$ field is related to the Lagrange multipliers 
 $\varepsilon$
 and $\mu$. Indeed, eliminating $p_\mu$ from the action \eqref{I1xp} we get
\be
\left. \hat{I}_1(x,p,\varepsilon,\mu)\right|_{p=\varepsilon^{-1}(\dot x- \mu x')} = 
\frac12 T\int d\tau d\sigma \left[ \varepsilon^{-1}(\dot x -\mu x')^2 - \varepsilon x'^2
\right].
\ee
Comparing the integrand here  with the one in the  action  with independent 2d metric 
$$
-\frac12 \sqrt{-g}g^{ab}\p_a x^\mu \p_b x_\mu =-\frac12\sqrt{-g} (g^{00} \dot x^2 +g^{11}
 x'^2+2 g^{01} \dot x x') ,
$$ 
we can identify 
\be
\varepsilon=-\frac{1}{\sqrt{-g}g^{00}}\ ,\ \ \  \qquad \mu=-\frac{g^{01}}{g^{00}}.
\ee
Thus one may say that 
the components of the  2-$d$ metric $g_{ab}$      play the role of the 
Lagrange multipliers for the  two  constraints.

The classical string motion described 
 by the Nambu-Goto action may  be interpreted as a motion in
  phase space subject to the two 
  non-linear constraints.
   Eliminating the 
   momenta $p_\mu$  leads to the   action with independent 2d  metric, while 
   solving also for  the Lagrange multipliers brings us back
    to the Nambu-Goto action.

By analogy with the particle mass shell constraint, the 
term $x'^2$ in the first  constraint $p^2 + x'^2=0$ 
can be interpreted as  an 
effective particle  mass. The second 
 constraint $p x'=0$  
 says that the string motion is transverse to the profile of the string.


In the gauge $\sqrt{-g}g^{ab}=\eta^{ab}$ the 
vanishing of the 2-$d$ scalar stress tensor
\be
T_{ab}=\p_a x^\mu\p_b x_\mu - \frac12 \eta_{ab} \eta^{cd}\p_c x^\mu \p_d x_\mu=0
\ee
means that
\be   h_{ab} \sim \eta_{ab} \ \rightarrow\  h_{00}+h_{11}=0\ , \qquad  h_{01}=0
\ee
Thus 
\be
T_{00}+T_{11}=0\ ,\ \ \ \ \  T_{01}=0 \quad \rightarrow \quad \dot x^2 +x'^2=0,\qquad 
 \dot x x'=0.
\ee 

\subsection{String equations in the orthogonal gauge}

Let us introduce the ``light-cone'' parametrization for the world-sheet
\be
\xi^\pm=\tau\pm\sigma \ , 
\ee 
\be
\p_\tau= \p_+ + \p_- \ , \qquad \p_\sigma= \p_+  - \p_-\ .
\ee
Then we get 
\begin{align}
\p_+\p_- x^\mu &= 0 \qquad \text{-- equation of motion} \label{me_pm} \\
\p_\pm x^\mu \p_\pm x_\mu &= 0 \qquad \text{-- constraints}
\end{align}
Consequently, we can easily write the general solution of \eqref{me_pm} as a sum 
of the left-moving  and right-moving waves with arbitrary profiles
\be
x^\mu=f^\mu_+(\xi^+) + f^\mu_-(\xi^-).
\ee 
The 
constraints  then have the form
\be
f'^2_+=0\ ,\ \  \qquad f'^2_-=0.
\ee

\subsection{String boundary conditions and some simple solutions}

In the conformal gauge the string action reads
\be
S=-\frac12 T \int d\tau d\sigma\ \p^a x^\mu \p_a x_\mu.
\ee
Its variation  has the form
\be
\delta S = -T \int d\tau d\sigma\ \delta x^\mu (-\p^2 x_\mu) - T\int d\tau d\sigma\
 \p^a (\delta x^\mu \p_a x_\mu),
\ee
where the 
variations of $x^\mu$ at the starting and the 
ending values of $\tau$  are 
 equal to zero, 
  $\delta x(\tau_{1},\sigma)=0$, $\delta x(\tau_{2},\sigma)=0$.

  \subsubsection{Open strings}
  
  If the string is open, 
   the 
    second term vanishes in the two cases:

\begin{minipage}[h]{0.4\linewidth}
\centering{\epsfig{figure=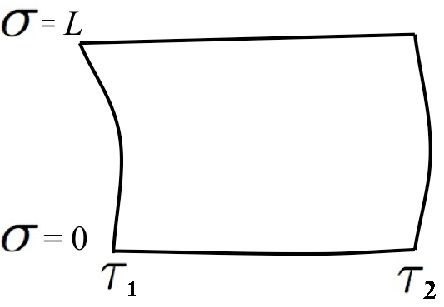,width=4cm,height=3cm}}

Fig.14
\vspace{0.3cm}
\end{minipage}
\hfill
\begin{minipage}[h]{0.5\linewidth}
{\it Neumann condition}: free ends of the string
\be
\left. \p_\sigma x^\mu(\tau,\sigma)\right|_{\sigma=0,L}=0
\ee
{\it Dirichlet condition}: fixed ends of the string
\be
\left. x^\mu(\tau,\sigma)\right|_{\sigma=0,L}=y_{0,L}^\mu(\tau)
\ee
\vspace{-0.5cm}
\end{minipage}
\noindent
Here $y_{0,L}^\mu(\tau)$ are some given trajectories.
The Dirichlet condition
is  relevant for the open-string 
description of $D$-branes.
 This  condition  breaks the Poincar\'{e}  invariance.

In general,  one can also impose 
mixed boundary conditions. Let us  divide the
components of $x^\mu$ into the two  sets

\begin{minipage}[h]{0.4\linewidth}
\centering{\epsfig{figure=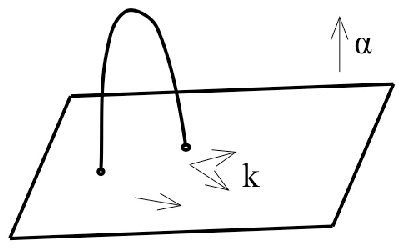,width=4cm,height=3cm}}

Fig.15: Boundary conditions
\vspace{0.3cm}
\end{minipage}
\hfill
\begin{minipage}[h]{0.5\linewidth}
Neumann components (along the brane)
$$
\{x^\alpha\} = \{x^0,x^1,\ldots,x^p\}
$$
Dirichlet components (transverse to brane)
$$
\{x^k\}=\{x^{p+1},x^{p+2},\ldots,x^{D-1}\}
$$
\vspace{-0.5cm}
\end{minipage}
The boundary conditions read
\be
\left. \p_\sigma x^\alpha\right|_{\sigma=0,L}=0\ , 
\qquad \left. x^k(\tau,\sigma)\right|_{\sigma=0,L}=y^k_{0,L}(\tau).
\ee
If the 
space-time contains no $Dp$-branes,
the  open strings have 
free ends in all the directions (Neumann conditions).  Then 
the full 
Poincar\'{e} invariance is unbroken.

\subsubsection{Closed strings}

For closed strings we impose the 
periodic condition $x^\mu(\tau,\sigma)=x^\mu(\tau,\sigma+L)$ and
may  choose units so that
$L=2\pi$ and $(\tau,\sigma)$ are dimensionless.
Then we get  the set of equations
\be
\ddot{x}-x''=0\ , \qquad \dot{x}^2+x'^2=0\ , \qquad \dot{x}x'=0
\ee
with the  condition
\be
x^\mu(\tau,\sigma)=x^\mu(\tau,\sigma+2\pi).
\ee
The simplest solution is a point-like string
\be
x^\mu(\tau,\sigma)=\bar{x}^\mu(\tau)=x_0^\mu+ p^\mu \tau\ ,
\ \ \ \ \ \  \ \ \ \ \ \
 p^\mu p_\mu=0\ . \ee

\subsection{Conservation laws}

It is useful to  split  $x^\mu(\tau,\sigma)$  into the coordinate 
 of the  center of mass
  $\bar{x}^\mu(\tau,\sigma)$ and oscillations around it $\tilde{x}^\mu(\tau,\sigma)$
\be \label{x-decomp}
x^\mu(\tau,\sigma) = \bar{x}^\mu(\tau) + \tilde{x}^\mu(\tau,\sigma),
\ee
\be \label{bar-x}
\bar{x}^\mu(\tau) = \frac{1}{L} \int d\sigma\ x^\mu(\tau,\sigma),
\ee
\be \label{tilde-x}
\int d\sigma\ \tilde{x}^\mu(\tau,\sigma) = 0.
\ee
Global symmetries lead via the Noether theorem to quantities that 
are   conserved  on the 
 equations of motion. 
 
 On the other hand, the local symmetries (reparametrizations and the Weyl transformation) lead to the 
 restrictions on the  energy-momentum tensor
\be
\p_a T^{ab}=0, \qquad T^a_a=0  \ ,  
\ee
or  to the constraints on $x^\mu$  after the gauge fixing. 

If the action is invariant under some transformation $\delta x^\mu$,  then using
Lagrange equations of motion
\be
\frac{\p \mathcal{L}}{\p x^\mu}-\p_a \frac{\p \mathcal{L}}{\p \p_a x^\mu}=0
\ee
we get a conserved current
\be \label{current}
j^a = \frac{\p \mathcal{L}}{\p \p_a x^\mu} \delta x^\mu, \ \ \ \qquad \p_a j^a=0.
\ee
Let $\delta x^\mu = \Lambda^\mu{}_A \varepsilon^A$, where $\varepsilon^A$ are constant 
parameters.
Then $j^a = j^a_A \varepsilon^A$ and $\p_a j^a_A=0$.  Now it is easy to check that  the 
integral
\be
J_A(\tau)=\int d\sigma j^0_A(\tau,\sigma)
\ee
gives a  conserved charge
\be
\frac{d}{d \tau} J_A(\tau) = 0.
\ee
The string action is invariant under the Poincar\'{e} transformations
\be 
\delta x^\mu= \varepsilon^\mu{}_\nu x^\nu +\varepsilon^\mu,
\ee
where $\varepsilon^\mu{}_\nu= -\varepsilon_\nu{}^\mu$, i.e. $\varepsilon_\mu{}^\nu \in so(1,D-1)$.
Then for space-time translations
in the  orthogonal gauge $g_{ab} \sim \eta_{ab}$ 
 we get the  conserved current (momentum density)
\be
 p^a = p^a_\mu \varepsilon^\mu =  \frac{\p \mathcal{L}}{\p \p_a x^\mu} \varepsilon^\mu= -T\p^a x_\mu\  
 \varepsilon^\mu.
\ee
Recalling that $\dot{x}=\p_0 x =-\p^0 x$ and taking \eqref{x-decomp} into account, 
we get 
\be \label{P-neter}
P^\mu = \int d\sigma\ p^{0\mu} =T\int d\sigma\ \dot{x}^\mu = TL\ \dot{\bar{x}}^\mu,
\ee
\be
\frac{d}{d\tau}P^\mu=0.
\ee
Indeed, using the equation of motion and the boundary conditions  we get 
\be
\dot{P}^\mu = T\int d\sigma\ \ddot{x}^\mu = T\int_{0}^{L} d\sigma\
 x''^\mu=T\left.(x'^\mu)\right|_{0}^{L}=0
\ee
This conservation law 
 means   that there is  no momentum flow through boundary.

From \eqref{P-neter} we get
\be
\bar{x}^\mu(\tau) = x^\mu_0 + p^\mu\tau, \qquad p^\mu=\alpha' P^\mu, \ \ \ \ \ \ \ \ \ \ 
TL \equiv {1 \over \alpha'} \ . 
\ee
Thus, the center of mass moves with a constant velocity  determined by the total 
momentum of the string.

\underline{Remark}:

Since the string equation of motion is of the second order formally  it has a simple solution linear in
both  $\tau$ and $\sigma$
\be \label{pq-sol}
x^\mu=x_0^\mu+p^\mu \tau +q^\mu \sigma.
\ee
For the open strings  with free ends, i.e. with  Neumann   boundary conditions  we get $q^\mu=0$.
In  the case of the closed strings  we  need 
 $x^\mu (\tau,\sigma)=x^\mu (\tau, \sigma+2\pi)$. A generalized version of this   condition  
 can be satisfied if $x^\mu $ is a compact (angular) coordinate, e.g., 

\vspace{0.3cm}
\begin{minipage}[h]{0.3\linewidth}
\centering{\epsfig{figure=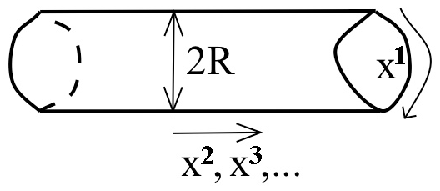,width=4cm,height=3cm}}

Fig.18: Compactification on cylinder
\vspace{0.6cm}
\end{minipage}
\hfill
\begin{minipage}[h]{0.5\linewidth}
$$
x^1=R\varphi , \qquad \varphi \sim \varphi+2\pi
$$
Then \eqref{pq-sol}  is consistent 
with  the ``generalized'' periodicity
$$
x(\tau, \sigma+2\pi)=x(\tau,\sigma) +2\pi R w,
$$
where $w=0,1,2,\ldots$ is the ``winding'' number.
\vspace{0.8cm}
\end{minipage}

\noindent
Then  $x=\bar{x}(\tau)+q\sigma$, where $q=Rw$ is  quantized ``winding momentum'' (cf. particle in a box).

Solution \eqref{pq-sol} has an additional symmetry (``$T$-duality'') under 
 $p \leftrightarrow q,\  \tau \leftrightarrow \sigma$.

\

Consider  next the Lorentz rotations $\delta x^\mu = \varepsilon^\mu{}_\nu x^\nu$.
The   conserved current (angular momentum density) reads
\be
j^a 
= \frac{\p \mathcal{L}}{\p \p_a x^\mu} \varepsilon^\mu{}_\nu x^\nu = \frac12 j^a_{\mu\nu} \varepsilon^{\mu\nu},
\ee
\be
j^{\mu\nu}_a(\tau,\sigma) = T(x^\mu\p_a x^\nu-x^\nu\p_a x^\mu)
\ee
The  corresponding conserved   charge  is 
\be \label{J-neter}
J^{\mu\nu} = \int d\sigma\ j^{0\mu\nu} =T\int d\sigma\ (\dot{x}^\mu x^\nu-\dot{x}^\nu x^\mu).
\ee
By construction, on the equations of motion one has 
\be
\frac{d}{d\tau}J_{\mu\nu}=0.
\ee
Using \eqref{x-decomp},  we can rewrite \eqref{J-neter} as
\be
J^{\mu\nu} = T\int d\sigma (\dot{\bar{x}}^\mu \bar{x}^\nu-\dot{\bar{x}}^\nu \bar{x}^\mu)
+ T\int d\sigma (\dot{\tilde{x}}^\mu \tilde{x}^\nu-\dot{\tilde{x}}^\nu \tilde{x}^\mu) = I^{\mu\nu}+S^{\mu\nu}.
\ee
Cross-terms here vanish due to \eqref{bar-x} and \eqref{tilde-x}. 
 $I^{\mu\nu}$ is the orbital angular momentum and $S^{\mu\nu}$ is
 the  internal one (i.e. the spin). Using \eqref{P-neter} we can rewrite  the orbital moment as
\be \label{orb-mom}
I^{\mu\nu} = P^\mu \bar{x}^\nu-P^\nu \bar{x}^\mu.
\ee 
Using  \eqref{P-neter} we have
$$
\frac{d}{d\tau}{I}^{\mu\nu} = \frac{d}{d \tau} (P^\mu \bar{x}^\nu-P^\nu \bar{x}^\mu)=
(P^\mu \dot{\bar{x}}^\nu-P^\nu \dot{\bar{x}}^\mu) = 0.
$$
With the help of the  equations of motion we get also 
\begin{align*}
\frac{d}{d\tau} {S}^{\mu\nu} &=  T\int d\sigma\ (\ddot{\tilde{x}}^\mu \tilde{x}^\nu-\ddot{\tilde{x}}^\nu \tilde{x}^\mu)
= T\int d\sigma\ (\tilde{x}''^\mu \tilde{x}^\nu-\tilde{x}''^\nu \tilde{x}^\mu) \\
&=T\int d\sigma\ \frac{\p}{\p \sigma}(\tilde{x}'^\mu \tilde{x}^\nu-\tilde{x}'^\nu \tilde{x}^\mu) 
=T (\tilde{x}'^\mu \tilde{x}^\nu-\tilde{x}'^\nu \tilde{x}^\mu\left.)\right|_0^L=0.
\end{align*}

  \def \ww {\omega}
  
\subsection{Rotating string solution}

One simple solution is provided by a folded closed 
string rotating in the 
 $(x^1,x^2)$-plane with its center of mass at rest 

\begin{minipage}[h]{0.4\linewidth}
\centering{\epsfig{figure=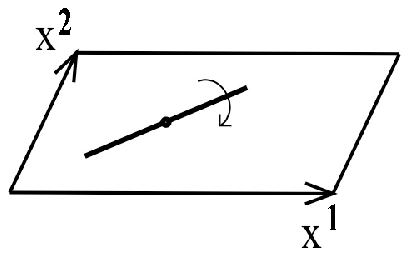,width=4cm,height=3cm}}

Fig.16: Rotating string
\vspace{0.1cm}
\end{minipage}
\hfill
\begin{minipage}[h]{0.5\linewidth}
$$
x^0=p^0 \tau
$$
$$
x^1=r(\sigma) \cos{\phi(\tau)}
$$
$$
x^2=r(\sigma) \sin{\phi(\tau)},
$$
$$
r(\phi)=a \sin{\ww \sigma},\ \ \ \ \ \ \ \  \phi(\tau)=\ww \tau
$$
\vspace{0cm}
\end{minipage}
\vspace{0.2cm}

\noindent
This  solves  $\ddot{x}-x''=0$  with $a$ being an arbitrary  constant.
 From the  periodicity condition it follows that
$w$ is integer, i.e. $\ww=0,\pm 1,\pm 2, \ldots$.
The constraint $\dot{x}^2+x'^2=0$ yields $p^0=a\ww$, and the 
 constraint $\dot{x}x'=0$ is satisfied automatically. 

The Lagrangian of the string is 
$$
\mathcal{L} = - \frac12 T \p_a x^\mu \p^a x_\mu
$$
so that the energy  related the  time-translation  symmetry $x^0 \rightarrow x^0+\epsilon $
is  (we fix $L=2 \pi$) 
\be \label{energy}
E = P^0= T\int_{0}^{2\pi} d\sigma\ \frac{\p \mathcal{L}}{\p \dot{x}^0}
= T\int_{0}^{2\pi} d\sigma\ \dot{x}^0 =\frac{p^0}{\alpha'} = \frac{a\ww}{\alpha'},
\ee
where  $\alpha'=\frac{1}{2\pi T}$. 

The spin is related to the  rotation symmetry $\varphi \rightarrow \varphi+\epsilon$
\be \label{spin}
S = T\int_{0}^{2\pi} d\sigma\ \frac{\p \mathcal{L}}{\p \varphi}
= T\int_{0}^{2\pi} d\sigma\ r^2(\sigma)\ \dot\varphi 
= Ta^2\ww\int_{0}^{2\pi} d\sigma\  \sin^2{\ww\sigma} 
=\frac{a^2\ww}{2\alpha'}.
\ee
Combining \eqref{energy} and \eqref{spin} we obtain  relation between the energy and the spin
\be \label{ESR}
\alpha' E^2 =2\ww S,\ \  \ \qquad w=1,2,\ldots
\ee
The configurations with the 
lowest energy for a given spin (or smalest slope on $E^2(S)$ plot) 
with $\ww=1$ belong to the leading Regge trajectory.
For them 
 $\alpha' E^2=2S$ or  recalling that $L=2\pi$ we have
\be
E^2=2TL S\  .
\ee
 
 \  

We can generalize the above rotating  solution
  by adding  motion to its center of mass:
\vspace{0.3cm}

\begin{minipage}[h]{0.4\linewidth}
\centering{\epsfig{figure=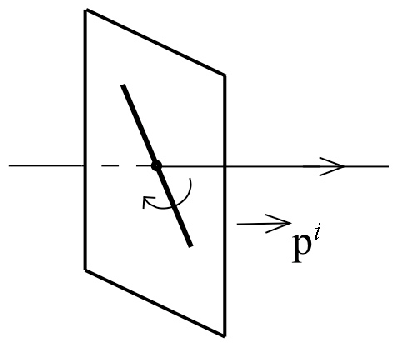,width=4cm,height=3cm}}

Fig.17: Rotating string with moving c.o.m.
\vspace{0.6cm}
\end{minipage}
\hfill
\begin{minipage}[h]{0.6\linewidth}

$$
x^0=p^0 \tau
$$
$$
x^1= x^1(\tau,\sigma), \qquad x^2= x^2(\tau,\sigma)
$$
$$
x^i = x^i_0+ p^i \tau, \qquad i=3,4,\ldots
$$
\vspace{0.5cm}
\end{minipage}

\noindent
The constraint $\dot{x}^2+x'^2=0$ then gives  $p_0^2-p_i^2=a^2\ww^2$. 
As in the case of the rotating string at rest we can find the expressions for the energy and 
the spin  \eqref{ESR}
\be
E^2 = P_i^2+\frac{2\ww}{\alpha'}S, \ \ \ \qquad P_i=\frac{p_i}{\alpha'}.
\ee


\subsection{Orthogonal gauge, conformal  reparametrizations and the light-cone gauge }

In the orthogonal gauge the string action has the form
\be
I = -      { 1 \over 2}  T\int d^2\xi \  \p_+ x^\mu \p_- x_\mu.
\ee
The equations of motion and the constraints read
\be
\p_+\p_- x^\mu = 0, \qquad \qquad \p_\pm x^\mu \p_\pm x_\mu = 0.
\ee
These are invariant under the 
residual conformal 
symmetry which is a subgroup of reparametrizations $\xi'^a = F_a(\xi^1,\xi^2)$
($\xi^\pm =  \tau \pm  \sigma$) 
\be
{\xi^+}' = F_+(\xi^+), \ \ \ \ \ \qquad {\xi^-}' = F_-(\xi^-).
\ee
Under such transformations the 2d  metric changes by a  conformal factor
\be
{ds^2}' =-d{\xi^+}'{d\xi^-}' = - \frac{dF_+}{d\xi^+}\frac{dF_-}{d\xi^-}d\xi^+d\xi^-
\ee
i.e. the  orthogonal-gauge condition is preserved by such conformal reparametrizations.


One can choose a special parametrization by using this additional invariance, 
i.e. on the equations of motion  for $x^\pm$  one can  impose an  additional
gauge condition  -- the  ``light-cone gauge''
\be
x^+(\tau,\sigma) = \bar{x}^+(\tau) = x_0^+ + p^+\tau  \ , \ \ \ \ \ \ \ \ \ \ \ 
 x^+ \equiv  x^0+x^{D-1}  \ . 
\ee 
That means that  $\tilde{x}^+(\tau,\sigma)=0$, i.e.
there is no oscillations in $x^+$.
 Indeed, since 
$$
\p_+\p_- x^+ = 0\qquad   \rightarrow\qquad  x^+ = f_1^+(\tau+\sigma) + f_2^-(\tau-\sigma),
$$
and one can use conformal reparametrizations to 
choose these functions so that 
$
 x^+ =\frac12 p (\tau+\sigma) + \frac12 p (\tau-\sigma) = p\tau,
$
where we also set $x_0^+=0$ by a shift. 

This is a  ``physical gauge'', in which only the transverse oscillations $x^i$ ($i=1,2,\ldots, D-2$)
 are dynamical.
Indeed, $\tilde{x}^-$ is then determined from the orthogonal gauge  constraints
\be
\p_\pm x^\mu \p_\pm x_\mu = 0 \ \ \ \rightarrow\ \ \ \ -\p_\pm x^+ \p_\pm x^- + \p_\pm x^i \p_\pm x^i = 0,
\ee
i.e. 
\be
\frac12 p^+ \p_\pm x^- = \p_\pm x^i \p_\pm x^i   \ , 
\ee
which is a linear equation on $x^-$ determining it in terms of $x^i$.

\vspace{0.3cm}
\begin{minipage}[h]{0.3\linewidth}
\centering{\epsfig{figure=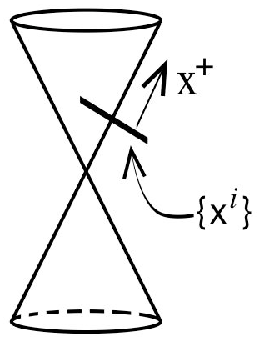,width=4cm,height=3cm}}

Fig.19: Light cone
\vspace{0.6cm}
\end{minipage}
\hfill
\begin{minipage}[h]{0.6\linewidth}
We end up with 
$$
\begin{cases}
x^+ = \bar{x}^+(\tau) \\
x^- = \bar{x}^-(\tau) + F[x^i(\tau,\sigma)] \\
x^i = \bar{x}^i(\tau) +\tilde{x}_L^i(\tau+\sigma) + \tilde{x}_R^i(\tau-\sigma)
\end{cases}
$$
$x^i$= dynamical degrees of freedom. 
\vspace{0.8cm}
\end{minipage}

This gauge  where $x^i$ are dynamical  is analogous to the 
light-cone gauge in  quantum electrodynamics ($A_+ = A_0+A_3 =0$)
where $A_i \ (i=1,2)$ are  dynamical degrees of freedom.

\section{General solution of string equations\\
 in the orthogonal gauge}


Let us use uniform  notation for open/closed case: $\sigma=(0,\pi),$ i.e. choose 
$  L=\pi$\\
Boundary conditions: 
\begin{align*}
& x(\tau,\sigma) = x(\tau,\sigma+\pi) &\text{closed string} \\
& \p_\sigma x \left.\right|_{\sigma=0} = \p_\sigma x \left.\right|_{\sigma=\pi} = 0 & \text{open string}
\end{align*}
Equations of motion  and constraints read
\be
\p_+\p_- x^\mu = 0,\ \ \ \ \  \qquad \p_\pm x^\mu \p_\pm x_\mu = 0.
\ee
We set  as before 
\be
x(\tau, \sigma) = x_0^\mu +p^\mu \tau +\tilde{x}^\mu(\tau,\sigma).
\ee
Then $ \tilde{x}^\mu$ is given by  Fourier mode expansion. 


\

\noindent {\bf Open string:}

Boundary conditions ${\tilde{x}'^\mu}\left.\right|_{\sigma=0,\pi}=0$
are satisfied by 
\be \label{sol-open}
\tilde{x}^\mu = \sum_{n \ne 0}^{\infty} a^\mu_n \,e^{-i n\tau} \cos n\sigma
\ee
Reality condition of $x^\mu$  implies: $(a^\mu_n)^*=a^\mu_{-n}$.
The momentum is 
\be
P^\mu =T\int_{0}^{\pi} d\sigma\ \dot{x}^\mu = T\pi {p}^\mu  = \frac{1}{2\alpha'} {p}^\mu
\ee
The angular momentum (=orbital+spin) is 
\be
J^{\mu\nu} =  I^{\mu\nu} + S^{\mu\nu} = \bar{x}^\mu p^\nu-\bar{x}^\nu p^\mu+
T\int_0^\pi d\sigma \, (\tilde{x}^\mu \dot{\tilde{x}}^\nu -\tilde{x}^\nu \dot{\tilde{x}}^\mu)
\ee
Rescaling  
$$a^\mu_n = i\frac{\sqrt{2\alpha'}}{n}\alpha^\mu_n$$
 we get
\be
\tilde{x}^\mu= i\sqrt{2\alpha'}\sum_{n \ne 0}^{\infty} \frac{\alpha^\mu_n}{n} \,e^{-i n\tau} \cos n\sigma.
\ee

\noindent {\bf Closed string:\\}

Here  $x(\sigma)=x(\sigma+\pi)$  and we can represent the oscillating part of the 
solution as a sum of the  left-moving and right-moving waves $
\tilde{x}^\mu(\tau,\sigma)
 =\tilde{x}_L^\mu(\xi^-) +\tilde{x}_R^\mu(\xi^+)
$ \ ($\xi^\pm = \tau \pm \sigma$), i.e. 
\be
{x}^\mu(\tau,\sigma)= x_R^\mu(\xi^-) + x_L^\mu(\xi^+)  \ , 
\ee
where 
\begin{align} \label{sol-closed}
x_R^\mu(\xi^-) &= 
\frac12 x_0^\mu + \frac12 \bar{p}^\mu \xi^- + \sum_{n \ne 0}^{\infty} a^\mu_n \,e^{-2in\xi^-} \\
x_L^\mu(\xi^+) &=
\frac12 x_0^\mu + \frac12 \bar{p}^\mu \xi^+ + \sum_{n \ne 0}^{\infty} \tilde{a}^\mu_n \,e^{-2in\xi^+} 
\end{align}
Reality condition implies: $(a^\mu_n)^*=a^\mu_{-n},\ (\tilde{a}^\mu_n)^*=\tilde{a}^\mu_{-n}$.

After the  rescaling $a \rightarrow i\sqrt{\frac{\alpha'}{2}}\frac{\alpha_n}{n}$,
\ $\tilde a \rightarrow i\sqrt{\frac{\alpha'}{2}}\frac{\tilde \alpha_n}{n}$,
 we get
\be
{x}^\mu= x_0^\mu + 2\alpha' P^\mu \tau + i\sqrt{\frac{\alpha'}{2}}\sum_{n 
\ne 0}^{\infty} \frac{\alpha^\mu_n}{n} \,e^{-2in(\tau-\sigma)} + 
i\sqrt{\frac{\alpha'}{2}}\sum_{n \ne 0}^{\infty} \frac{\tilde{\alpha}^\mu_n}{n} \,e^{-2in(\tau+\sigma)}
\ee
The constraints imply the vanishing 
of the  energy-momentum tensor $T_{\pm\pm} = (\p_\pm x)^2=0$.
Let us introduce  its  Fourier  components. For  closed string  we get 
\begin{align}
L_n &= \frac{T}{2} \int d\xi^-\, e^{-2in\xi^-} T_{--}(\xi^-) \\
\bar{L}_n &= \frac{T}{2} \int d\xi^+\, e^{-2in\xi^+} T_{++}(\xi^+) 
\end{align} 
where  we integrate over $\sigma$ in the range $(0,\pi)$ at  $\tau=0$.
For the open string
\be
L_n = T \int_0^\pi d\sigma^- \, (e^{in\sigma} T_{++} + e^{-in\sigma} T_{--})
=\frac{T}{4} \int_{-\pi}^\pi d\sigma\, e^{in\sigma} (\dot{x}+x')^2.
\ee
The  substitution of the solutions  \eqref{sol-open} and \eqref{sol-closed} into 
 these  integrals yields
\begin{align} \label{L-closed}
L_n = \frac12 \sum_{k=-\infty}^{\infty} \alpha^\mu_{n-k} \alpha_{k\mu}, \  \ \ \ \ \ \ \ \ \ \ \ \  \ \ \ 
\bar{L}_n = \frac12 \sum_{k=-\infty}^{\infty} \tilde{\alpha}^\mu_{n-k} \tilde{\alpha}_{k\mu}
\end{align}
for the closed string and
\be \label{L-open}
L_n= \frac12 \sum_{k=-\infty}^{\infty} \alpha^\mu_{n-k} \alpha_{k\mu}
\ee
for the  open string.
Here we introduced the following definition:
\begin{align} \label{alpha0}
\tilde{\alpha}_0^\mu &=\alpha_0^\mu = \sqrt{\frac{\alpha'}{2}}p^\mu &\text{closed string}\\
\alpha_0^\mu &= \sqrt{2\alpha'}p^\mu &\text{open string}
\end{align}

 The angular momentum is found to be 
\be 
J^{\mu\nu} = I^{\mu\nu} + S^{\mu\nu}(\alpha) + S^{\mu\nu}(\tilde{\alpha}),
\ee
where
$$
I^{\mu\nu} = \bar{x}^\mu p^\nu- \bar{x}^\nu p^\mu,
$$
$$
S^{\mu\nu}(\alpha)  = i\sum_{n=1}^{\infty}\frac{1}{n}(\alpha^\mu_{-n}\alpha^\nu_n - \alpha^\nu_{-n}\alpha^\mu_n)
$$

The string  Hamiltonian  is 
\be
H= T\int_0^\pi d\sigma\, (\dot{x} p -\mathcal{L}) = \frac{T}{2} \int_0^\pi d\sigma\, (\dot{x}^2+{x'}^2)
\ee
i.e. 
\begin{align} \label{hamil}
H&=\frac12 \sum_{n=-\infty}^{\infty}\alpha^\mu_{-n}\alpha_{n\mu} =L_0  & \text{open string}\\
H&=\frac12 \sum_{n=-\infty}^{\infty}(\alpha^\mu_{-n}\alpha_{n\mu}+\tilde{\alpha}^\mu_{-n}\tilde{\alpha}_{n\mu}) =L_0 +\bar{L}_0  & \text{closed string}
\end{align}
 From \eqref{alpha0}, \eqref{hamil} and
mass condition $p^2=-M^2$ it follows that
\begin{align}
\alpha' M^2 &= \sum_{n=1}^{\infty}\alpha^\mu_{-n}\alpha_{n\mu}  & \text{open string}\\
\alpha' M^2 &= 2\sum_{n=1}^{\infty}(\alpha^\mu_{-n}\alpha_{n\mu}+\tilde{\alpha}^\mu_{-n}\tilde{\alpha}_{n\mu})   & \text{closed string}
\end{align}


Let us introduce the Poisson brackets 
\begin{align}
\{x^\mu(\sigma,\tau),x^\nu(\sigma',\tau)\} &= \{\dot{x}^\mu,\dot{x}^\nu\} = 0, \\
\{\dot{x}^\mu(\sigma,\tau),x^\nu(\sigma',\tau)\} &= T^{-1} \delta(\sigma-\sigma')\eta^{\mu\nu},
\end{align}
Using the $\delta$-function  representation
$$
\delta({\sigma}) = \frac{1}{\pi} \sum_{n=-\infty}^{\infty} e^{2in\sigma}
$$
one can check that
\begin{align}
\{\alpha^\mu_m, \alpha^\nu_n\} &= \{\tilde{\alpha}^\mu_m, \tilde{\alpha}^\nu_n\} = 
im\delta_{m+n,0}\eta^{\mu\nu}, \\
\{\alpha^\mu_m, \tilde{\alpha}^\nu_n\} &=0.
\end{align}
For the  zero modes  we get $\{p^\mu, \bar{x}^\nu\} = \eta^{\mu\nu}$.

\ 

One finds then that 
\be
\{L_m,L_n\} = i(m-n)L_{m+n}   \ , 
\ee
\begin{align}
\{\bar{L}_m, \bar{L}_n\} = i(m-n)\bar{L}_{m+n},  \  \ \ \ \ \ \ \ \ \ \ \ \ 
\{L_m, \bar{L}_n\} = 0 \ , 
\end{align}
i.e. the constraints are in involution forming the Virasoro algebra. 

From \eqref{L-closed} and \eqref{L-open} it follows that
\be
(L_n)^* = L_{-n}, \ \ \ \qquad (\bar{L}_n)^* = \bar{L}_{-n}.
\ee

\section{ Covariant quantization of free string
}

\subsection{Approaches to quantization}

I. {\bf Canonical operator approach}

Here  one starts with promoting classical canonical variables to operators
and imposing standard commutation relations. 

In the Lorentz-covariant   approach
one uses 
covariant  orthogonal gauge  and imposes  constraints on states.
The Weyl symmetry is broken at the quantum level unless dimension 
$D=D_{crit} = 26$ (or $D=10$ for superstrings). 
 In this approach the theory is manifestly Lorentz-invariant, 
 but there are   no ghosts and the theory is unitary only  in
  $D=D_{crit}$.

In the  light cone  (``physical'')  gauge approach 
the 
constraints are first solved explicitly at the classical level, 
and then only the remaining transverse modes are quantized.
Here the theory is  manifestly unitary but is Lorentz-invariant only  
 in
  $D=D_{crit}$.

\

\

\noindent 
II. {\bf Path integral approach}

Here one starts    with the  path integral  defined by the string action 
$$
<\ldots> = \int Dg_{ab}\, \int Dx^\mu\, e^{i S[x,g]}\  \ldots
$$
%
The 2d 
metric is assumed to be non-dynamical
(the Weyl symmetry is assumed to be an exact symmetry of the quantum theory, which happens to be true only if  
$D=D_{crit}$),  
 and the  integral over the continuous (local) modes  of the metric  is removed by the gauge fixing.

The observables  here 
are correlation functions of ``vertex operators'' (corresponding to string states)
 defining the  scattering amplitudes of particular string modes.

\subsection{Covariant operator  quantization}

 We promote 
canonical center-of-mass variables $(x_0,p)$ to operators  $(\hat x_0,\hat p)$, and do  the same for the  
 oscillation modes:  $\alpha \ra \hat\alpha, \tilde\alpha \ra \hat{\bar\alpha}$.\\
The Poisson brackets $\{\,,\,\} $  are replaced by the quantum commutator $-i[\,,\,]$.\\
The string state $(p, \alpha,\tilde\alpha) $  then  corresponds 
to a quantum state vector $|\psi>$ in Hilbert space.\\
The reality condition 
$\alpha_n=\alpha_{-n}^* $ becomes  $ \hat \alpha_n=\hat \alpha_{-n}^\dagger$.

 Then we get 
$[\hat x_0^\mu, \hat p^\nu]=i\eta^{\mu\nu}$, where
$\hat x^\mu =\hat  x_0^\mu +2\alpha' \hat p^\mu + \ldots$ and also 
(below we shall often omit  ``hats'' on the operators)
\begin{align}
[\alpha_m^\mu,\alpha_{-m}^\nu] = m\,\eta^{\mu\nu} \ , \qquad \qquad 
[\alpha_m^\mu,\alpha_n^\nu] = 0, \qquad m \ne -n
\end{align}
The  ``Fock vacuum'' $|0,p>$  satisfies 
\be
\hat{p}|0,p> = p|0,p>.
\ee
The vacuum is annihilated by $\alpha_m^\mu, \  m>0$
\be
a_m^\mu |0,p> = 0 \quad  (m>0) \ ,\ \ \ \ \ \ \ \ \ \ \ 
a_m^\mu \equiv  \frac{1}{\sqrt{m}}\alpha_m^\mu
\ee 
The general string 
state vector is
\be
|\psi> = (a_{n_1}^\dagger )^{k_1} \ldots (a_{n_l}^\dagger )^{k_l}\,|0,p>
\ee
Not all states are physical. They must satisfy the constraints and have positive norm.
But 
$$
 [\alpha_n^0,\alpha_n^{0\dagger}] = -1 \ ,  \qquad n>0,
$$
implies existence of negative norm states
$$
||\alpha^0_n\,|0>|| = -1
$$
These  can be ruled out by imposing the quantum version of the constraints.

\subsubsection{Open string}

Here the  Virasoro operators are 
\begin{align}
L_0 &= \alpha' p^2 + \sum_{n=1}^{\infty} \alpha^\mu_{-n} \alpha_{n\mu}\\
L_m &= L_{-m}^\dagger = \frac12 \sum_{n=0}^{\infty} \alpha^\mu_{m-n} \alpha_{n\mu}\ , \qquad m>0.
\end{align}
In the classical theory the constraints are 
$L_0=0,\  L_m = 0$. In the quantum theory $L_0$ and $L_m$ become
operators and the problem of ordering of  the operators that 
enter them  arises.

The normal ordering is automatic for $m \ne 0$ since $m-n \ne -n$
\be
L_m \rightarrow \hat{L}_m = \frac12 \sum_{n=0}^{\infty} \alpha_{m-n}^\mu\alpha_{n\mu} \ , \qquad m \ne 0.
\ee
For $m=0$ we have in general ($c_1+c_2=1$)
\begin{align}
L_0 \rightarrow \hat{L}_0 &= \alpha'p^2+  \sum_{n=1}^{\infty} (c_1\, \alpha_{-n}^\mu\alpha_{n\mu}+
c_2\, \alpha_{n}^\mu\alpha_{-n\mu})\notag \\
&= \alpha'p^2 + \sum_{n=1}^{\infty}  \alpha_{-n}^\mu\alpha_{n\mu} +
c_2\, \sum_{n=1}^{\infty} [\alpha_{n}^\mu, \alpha_{-n\mu}] \notag
\end{align}
 Then 
\be
\hat{L}_0 = :\hat{L}_0: + a\ ,\qquad \qquad a=c_2\, D\sum_{n=1}^{\infty} n.
\ee
It turns out that for  consistency of the theory 
one is to choose $a=-\frac{D}{24}$. This value 
 is found if one uses the  $\zeta$-function regularisation and 
 chooses also  $c_1=c_2=\frac12$.
 Consider the series 
$$
I_0(\varepsilon) = \sum_{n=1}^{\infty} e^{-\varepsilon n} = \frac{1}{e^{\varepsilon}-1}
\xrightarrow{\varepsilon \rightarrow 0}\ \  \frac{1}{\varepsilon} - \frac12 + O(\varepsilon)
$$
$$
I_1(\varepsilon) = - \frac{d}{d\varepsilon}I_0=  \sum_{n=1}^{\infty} ne^{-\varepsilon n} = \frac{e^{\varepsilon}}{(e^{\varepsilon}-1)^2}
\xrightarrow{\varepsilon \rightarrow 0}\ \  \frac{1}{\varepsilon^2} - \frac{1}{12} + O(\varepsilon)
$$
$$
I_{-1}(\varepsilon) = - \int d\varepsilon I_0(\varepsilon) =  \sum_{n=1}^{\infty} \frac{1}{n}e^{-\varepsilon n} = -\ln(1-e^{-\varepsilon})
\xrightarrow{\varepsilon \rightarrow 0}\ \  -\ln{\varepsilon} + O(\varepsilon)
$$
The  $\zeta$-function is defined by 
\be
\zeta(s) =  \sum_{n=1}^{\infty} \frac{1}{n^s} ,  \ \ \ \ \ \ \qquad s \in \mathbb{C}
\ee
and then at  $s=0,1,-1, \ldots$ by an analytic continuation.  This gives 
the same values as finite parts in the above  series 
\be
\zeta(0) = -\frac12, \ \qquad \zeta(-1) = -\frac{1}{12},\  \qquad \zeta(1) = \infty
\ee
With this prescription 
\be
a= \frac12 D \zeta(-1) = -\frac{D}{24} \ .
\ee
In fact, the true  value of the normal 
 ordering constant $a$ is not $-\frac{D}{24}$ but $a=-\frac{D-2}{24}$:
 the  additional -2  contribution to $D$ 
 comes from the ``ghosts'' of the conformal gauge 
 (in the light-cone gauge, only  $D-2$ transverse modes $x^i$ contribute).
 
 \
 
The quantum version of the   constraints is then 
\begin{eqnarray}
(:L_0: +a) |\psi> & =& 0, \\
L_m |\psi> &=& 0, \qquad m>0
\end{eqnarray}
where
\begin{eqnarray}
:L_0: = \alpha' p^2 + \sum_{n=1}^{\infty} \alpha^\mu_{-n} \alpha_{n\mu}, \qquad \ \ \ \ \ 
L_m = \sum_{n=0}^{\infty} \alpha^\mu_{m-n} \alpha_{n\mu}
\end{eqnarray}
The mass shell condition $p^2=-M^2$  then  gives the expression for the mass operator
\be
\alpha'M^2 |\psi> = (a + \sum_{n=1}^{\infty}\alpha^\mu_{-n} \alpha_{n\mu}) |\psi>
\ee
In particular, the value $a$ thus determines the  mass of the ground state 
\be
\alpha'M^2 |0,p> = a |0,p> \ , 
\ee
which is tachyonic for $D=26$:\  $a= -\frac{D-2}{24}=-1$. 
In  the supersymmetric string  case  the ground state 
turns out  to be   massless.

\

For the Virasoro algebra one finds 
\begin{eqnarray}
\text{Classical:}& \qquad &\{L_m,L_n\}  = (m-n) L_{m+n}\\
\text{Quantum:}& \qquad &[L_m,L_n]  = (m-n) L_{m+n} + \frac{D}{12} (m^2-1)m\,\delta_{m+n,0}
\end{eqnarray}
where the quantum algebra turns out to contains a central  term.
When $L_m$ are modified by the ghost contributions 
the   coefficient $D$  in the central term is shifted to $D-2$.

Note that for  $n=-m$ we get 
\be
[L_m,L_{-m}] = 2mL_0 +\frac{D}{12}(m^3-m)
\ee
so that  in view of $(L_0+a)|\psi> = 0$ 
\be
[L_m,L_{-m}] |\psi> = 2m\Big[L_0 +\frac{D}{24}(m^2-1)\Big]|\psi> \ \ne \ 0,
\ee
which is in contradiction with 
 $L_m |\psi> =0$ for $m \ne 0$.
To avoid this contradiction it is sufficient to impose 
the constraints ``on average'', i.e.  to demand that matrix elements   of $L_m$ with  $m \ne 0$
should vanish:
\be
<\psi'|L_m|\psi> = 0 \ , \ \ \ \ \ \ \ m \ne 0
\ee  
The  spectrum is then described in terms of the oscillator states: 
\be
\hat{L}_0 = L_0 + a = \alpha' p^2 + \sum_{n=1}^{\infty}nN_n + a,
\ee
\be
N_n \equiv  \frac{1}{n} \alpha^\dagger_{\mu n}\alpha^\mu _n = a^\dagger_{\mu n} a^\mu_n,\  \ \    
\qquad [a^\mu _n,a^{\nu \dagger} _n] = \eta^{\mu\nu}  \ . 
\ee
$N_n$ has the following properties
\be
N_n |0> = 0, \qquad  N_n a^{\nu \dagger} _n |0> = a^{\nu \dagger} _n |0>.
\ee
For generic oscillator state 
\be
|\psi> = (a^\dagger _{n_1})^{i_1} \ldots (a^\dagger _{n_k})^{i_k} |0,p> \ , \ \ \ \ \ \ \ \ \ \ \ \ \ 
\hat p^\mu |0,p> = p^\mu |0,p> 
\ee
where  $(a^\dagger _{n_1})^{i_1}$ stands for  $  a^{\mu_1 \dagger} _{n_1} ...  a^{\mu_{i_1} \dagger}_{n_1}
 $
   we can define  the 
level  number  as 
\be
\ell= i_1 n_1 + \ldots + i_k n_k 
\ee
Then 
\be
\sum_{n=1}^{\infty} nN_n |\psi> \  \  = \  ( i_1 n_1 + \ldots + i_k n_k )|\psi> = \  \ell |\psi> 
\ee
and thus the mass  of this state is determined by 
 ($T= { 1 \over 2 \pi \alpha'}$) 
\be
M^2 |\psi>  =    \frac{1}{\alpha'}\Big(\sum_{n=1}^{\infty} nN_n+a\Big)|\psi>\   = \    2\pi T (\ell+a) |\psi>.
\ee

In the open string case we may call the physical states those that satisfy the condition 
\be
L_m |\psi> = 0 \qquad \text{for } m \ge 1.
\ee
Let us define the ``true physical states'' as those   that  have positive norm
\be
<\psi|\psi> \ \  > 0.
\ee
Other physical  states that have  zero norm are called  ``null states''.
One can prove the  
 ``no ghost'' theorem:  iff $D=26$  all physical states have 
 non-negative norm. 

\vspace{0.3cm}
\begin{minipage}[h]{0.3\linewidth}
\centering{\epsfig{figure=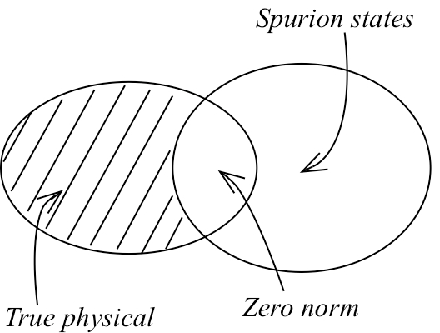,width=4cm,height=3cm}}

Fig.20: String states 
\vspace{0.6cm}
\end{minipage}
\hfill
\begin{minipage}[h]{0.6\linewidth}

One can define spurion states  as   those obtained by acting by $L_{-m}$, 
i.e.  $|\chi> = L_{-m}|\phi>$ for $m>0$.
Such states orthogonal to  physical since
$$
L_m|\psi>=0 \quad \rightarrow \quad  <\psi|L^\dagger _{m}=<\psi|L_{-m}=0 
$$
$$
<\psi|L_{-m}|\phi>=0 \quad\rightarrow\quad <\psi|\chi>=0.
$$
\vspace{0.8cm}
\end{minipage}
Null states   may be both physical and spurion; true physical states 
are then non-spurion ones, i.e. 
physical ones that have positive norm.

\subsubsection{Closed string}

Here we get two independent sets of creation-annihilation operators 
$\alpha^\mu_m$  and $ \tilde \alpha^\mu_m$.
The normal ordering ambiguity in going from classical to quantum theory implies 
$$
L_0 \rightarrow L_0 +a, \qquad \bar{L}_0 \rightarrow \bar{L}_0 +a.
$$
\be
\hat{L}_0+\hat{\bar{L}}_0 = \frac{\alpha'p^2}{2} + 
\sum_{n=1}^{\infty} (\alpha^\mu_{-n} \alpha_{n\mu} + \tilde\alpha^\mu_{-n} \tilde\alpha_{n\mu}) +2a,
\ee
\begin{eqnarray}
\hat{L}_m = \sum_{n=-\infty}^{\infty}  \alpha^\mu_{m-n} \alpha_{n\mu},\  \ \ \ \ \ \ \ \ \ \ \ \ \ \ \ 
\hat{\bar{L}}_m = \sum_{n=-\infty}^{\infty}  \tilde\alpha^\mu_{m-n} \tilde\alpha_{n\mu}
\end{eqnarray}
\be
 a_{closed} =2a_{open}=2a  = -\frac{D }{12} , 
\ee
where $D$ is shifted to $D-2$ after one accounts for the conformal gauge ghosts
or uses the light-cone gauge.

Here one gets  two copies of the Virasoro   algebra -- for $L_n$ and $\bar L_n$. 

The vacuum state   is defined by 
\be \hat p^\mu |0,p> = p^\mu |0,p>  \ , \ \ \ \ \ \ \ \ \ \ \ \
\alpha^\mu_n |0,p> = 0, \qquad \tilde\alpha^\mu_n |0,p> = 0 \ , \qquad n>0
\ee
and 
\be
|\psi> = (\alpha^\dagger _{n_1})^{i_1} \ldots (\alpha^\dagger _{n_k})^{i_k} 
\  (\tilde\alpha^\dagger _{m_1})^{j_1} \ldots (\tilde\alpha^\dagger _{m_s})^{j_s}  |0>
\ee
form an infinite set of states in Fock space.
The level numbers  are defined  by 
\be
\ell =i_1n_1+\ldots + i_kn_k\ , \qquad \bar{\ell } = m_1 j_1+\ldots + m_s j_s.
\ee


The constraints 
in  classical theory  are 
$
L_0=0, \bar{L}_0=0$ and $ L_m=0, \bar{L}_m=0 $ for $m \ne 0.
$
In quantum theory we need to impose them ``on average''  to avoid contradiction with the centrally extended
Virasoro algebra;  that means we need to assume that 
\begin{align}
(L_0+a)|\psi> &= 0,  \quad (\bar{L}_0+a)|\psi> = 0,\label{L0} \\
L_m |\psi> &= 0, \quad \bar{L}_m|\psi> = 0 \qquad \text{for } m>0.
\end{align}
Equivalently, \eqref{L0} reads
\begin{align}
(L_0+\bar{L}_0+2a)|\psi> &= 0,\label{L0p}\\
(L_0-\bar{L}_0) |\psi> &= 0.\label{L0m}
\end{align}
Then the constraints are satisfied for the expectation values
\begin{align}
<\psi'|L_m|\psi> = 0, \qquad <\psi'|\bar{L}_m|\psi> = 0, \qquad   m \ne 0.
\end{align}
Equation \eqref{L0p} determines the mass spectrum and \eqref{L0m} is 
the ``level matching'' condition.

From \eqref{L0p} it follows that
\be
\Big(\frac{\alpha'p^2}{2}+\sum_{n=1}^{\infty}n(N_n+\tilde{N}_n)+2a\Big)|\psi>=0,
\ee
where
\be
N_n = a^{\mu \dagger} _n a_{\mu n} = \frac{1}{n}\alpha^{\mu \dagger} _n \alpha_{\mu n}\ 
, \qquad \tilde{N}_n = \tilde{a}^{\mu \dagger} _n \tilde{a}_{\mu n}
= \frac{1}{n}\tilde\alpha^{\mu \dagger} _n \tilde\alpha_{\mu n} \  . 
\ee
Then the closed-string states   have masses 
\be
M^2 =\frac{2}{\alpha'}(\ell+\tilde{\ell}+2a).
\ee
Eq. \eqref{L0m}  gives 
\be
\sum_{n=1}^{\infty}n(N_n-\tilde{N}_n)|\psi>=0, \ \ \ \ \ {\rm i.e.} \ \ \ \ \ 
\ell= \tilde \ell
\ee
so that  we get 
\be
M^2 = \frac{4}{\alpha'}(\ell+a).
\ee
The mass of the ground state is thus 
\be
M^2_0 = \frac{4}{\alpha'} a \ .
\ee
As in the  open string case with $D=26$ here  we get 
$ 2a= -\frac{D-2 }{12}= -2$, so  that the ground state is tachyonic. 

The structure of the physical state space here is similar to the one 
in the open string case.


\section{Light cone gauge description of the free string spectrum}

The equations of motion and the constraints in the orthogonal gauge 
\be
\ddot{x}-x''=0, \qquad \dot{x}^2+x'^2=0, \qquad \dot{x}x'=0.
\ee
have residual  conformal reparametrization symmetry
\be
\xi^+ \rightarrow f(\xi^+), \qquad \ \ \  \xi^- \rightarrow \tilde{f}(\xi^-),\ \  \qquad 
\sigma^\pm = \tau\pm \sigma
\ee
As was already mentioned above, it  can be  fixed by imposing an additional 
``light cone gauge'' condition 
\be
x^+(\sigma,\tau) = \bar{x}^+(\tau) = x^+_0+2\alpha'p^+\tau, \qquad \ \ \ 
     x^\pm = x^0 \pm x^{D-1}, 
\ee
implying that  there is no oscillator part in the classical solution for    $x^+$.

Then the constraints  determine 
\be
x^-(\sigma,\tau) = x^-_0 +2\alpha'p^-\tau +F(p^+, x^i(\sigma,\tau)).
\ee
in terms of  $x^i$, i.e.  $D-2$ independent ``transverse'' degrees of freedom.

\subsection{Open string}

Using constraint $\dot{x}\cdot x'=0$ we get 
\be
0 = \dot{x}^\mu x'_\mu = -\frac12(\dot{x}^+x'^- + \dot{x}^-x'^+) +\dot{x}^ix'^i = 
-\alpha'p^+ +\dot{x}^ix'^i,
\ee
so that 
\be
x^- = \frac{1}{\alpha'p^+}\int d\sigma\ \dot{x}^ix'^i + h(\tau)
\ee
and thus 
\be
\alpha^-_n = \frac{1}{\sqrt{2\alpha'}p^+} \sum_{m=0}^{\infty}\alpha^i_{n-m}\alpha^i_{m}.
\ee
From $\dot{x}^2+x'^2=0$ it follows that
\be
2\alpha'p^+ \dot{x}^- = \dot{x}^i\dot{x}^i + x'^i x'^i.
\ee
Here we quantize as independent ones   the transverse oscillators only. 
The mass shell condition 
 $-\alpha'p^2 = \alpha' M^2 = N= N_\perp$ 
is then determined by the transverse-mode oscillation number 
\be
N_{\perp} = \sum_{n=1}^{\infty}\alpha^i_{-n}\alpha^i_{n}.
\ee
In quantum theory 
\be
\hat{N}_{\perp} =\, :N_{\perp}:\  +\  a\ , \ \ \ \ \ \ \ \ \ \ \ \ a = -\frac{D-2}{24}
\ee
If $D=26$ then $a=-1$ and one can show that the resulting 
spectrum is consistent with the requirement of the  Lorentz invariance
of the theory.

Since the constraints are solved already, 
by acting on the Fock vacuum by the creation operators $\alpha^i_{-n} = ( \alpha^i_{n})^\dagger$
we get only the physical states
\be
|\psi> = \xi_{i_1 \ldots i_m} (p) \alpha^{i_1}_{-n_1}\ldots \alpha^{i_m}_{-n_m}|0,p>
\ee
\be
\alpha^{i}_{n>0}|0,p>=0 \ , \ \ \ \ \ \ \  \ \ \ \ 
 \hat p^\mu |0,p> = p^\mu |0,p>,\ \ \ \ \  \qquad [x_0^\mu, p^\nu] = i\eta^{\mu\nu}
\ee
The angular momentum   $J^{\mu\nu} = I^{\mu\nu}+S^{\mu\nu}$  here is 
\be
I^{\mu\nu} =x^\mu p^\nu-x^\nu p^\mu, \qquad S^{\mu\nu} = -2i\alpha' 
\sum_n (\alpha^\mu_{-n}\alpha^\nu_{n}- \alpha^\nu_{-n}\alpha^\mu_{n}).
\ee
The Lorentz algebra is defined by
\be
[J^{\mu\nu}, J^{\lambda\rho}] = \eta^{\mu\lambda}J^{\nu\rho} - \eta^{\nu\lambda}J^{\mu\rho}-
\eta^{\mu\rho}J^{\nu\lambda} + \eta^{\nu\rho}J^{\mu\lambda},
\ee
where
$$
J^{\mu\nu} = (J^{ij}, J^{+i}, J^{-i}, J^{+-}).
$$
The commutation relation $[J^{-i},J^{-j}]=0$ realized
 in terms of the  quantum string oscillators turns out to be  valid  only if
$D=26$: the light cone gauge preserves Lorentz symmetry iff 
$D=26$ (in superstring theory one finds this for   $D=10$).

The lowest-mass states in the spectrum are: 

\vspace{0.5cm}
$N_{\perp}=\ell=0: \qquad \alpha'M^2=a=-1$

\vspace{0.5cm}
$N_{\perp}=\ell=1: \qquad \alpha'M^2=1-1=0$
$$
|\psi> = \xi_i(p)\alpha^i_{-1}|0,p>, \qquad i=1,\ldots, D-2
$$
$\xi_i(p)$ is a vector of $SO(D-2)$ having $D-2=24$ physical polarisations;
this  is consistent with Lorentz invariance since it is  massless.

\vspace{0.5cm}
$N_{\perp}=\ell=2: \qquad \alpha'M^2=2-1=1$
\noindent
There are two possibilities
\begin{align}
|\psi>&= \xi_i(p)\alpha^i_{-2}|0,p>\\
|\psi>' &= \xi_{ij}(p)\alpha^i_{-1}\alpha^j_{-1}|0,p>
\end{align}
The total number of components is
$$
D-2 +\frac{(D-2)(D-2+1)}{2} =\frac{D(D-1)}{2}-1.
$$
This is a dimension of $SO(D-1)$   representation (symmetric traceless tensor of $SO(D-1)$)
as it should be for 
massive state in a Lorentz-invariant theory in $D$ dimensions. 
Thus the  $\ell=2$ state is a spin-2 massive particle.

Higher levels are described by  $SO(D-1)$ representations as well.
Let us recall that the 
irreducible representations of $SO(r)$
are described by 
tensors $
t_{m_1\ldots m_k}, \ \  m_i=1,\ldots, r$
which are 
symmetric traceless, or 
 antisymmetric
or have mixed symmetry   and 
 can be represented  by the Young tableaux.

 Higher level  states in the 
 light cone gauge spectrum 
  are described by tensors of $SO(D-2)$  which 
   combine into  irreducible representations of $SO(D-1)$, 
   i.e.  by  massive particles in $D$ dimensions.

\vspace{0.3cm}
The maximal-spin state at  level  $\ell$  with  $\quad \alpha'M^2 =l-1$
  is 
\be
|\psi> = \xi_{i_1\ldots i_l}(p)\ \alpha^{i_1}_{-1}\ldots \alpha^{i_l}_{-1}|0,p>,
\ee
where $\xi_{i_1\ldots i_l}$ is a symmetric tensor of $SO(D-2)$. 
To get the full  state of spin $\ell $ one has to add lower  tensors of   $SO(D-2)$.

For instance, for $\ell=3$:
\begin{align*}
&\xi_{ijk} \alpha^{i}_{-1}\alpha^{j}_{-1}\alpha^{k}_{-1}|0,p> \qquad \text{symmetric 3rd rank tensor  }\\
&\xi_{ij} \alpha^{i}_{-2}\alpha^{j}_{-1}|0,p> \qquad \ \ \ \ \text{symmetric + antisymmetric 2nd rank tensor }\\
&\xi_{i}\alpha^{i}_{-3}|0,p> \qquad\ \ \ \ \ \ \ \ \ \text{vector}
\end{align*}

\vspace{0.3cm}
\begin{minipage}[h]{0.3\linewidth}
\centering{\epsfig{figure=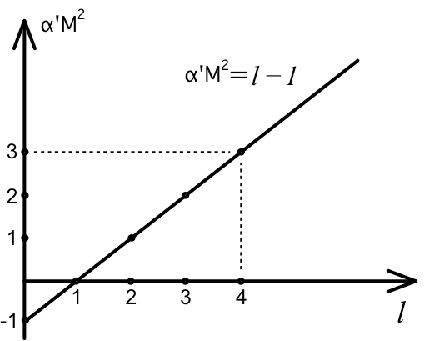,width=6cm,height=4cm}}

Fig.21: Regge trajectory
\vspace{0.6cm}
\end{minipage}
\hfill
\begin{minipage}[h]{0.5\linewidth}
Leading Regge trajectory  (highest spin states at given level):
$$
J=\ell= \alpha(E) = \alpha(0) + \alpha' E^2 = 1+\alpha'M^2
$$
\vspace{0.8cm}
\end{minipage}

\subsection{Closed string}

Here the transverse oscillators are $(\alpha^i_n, \tilde{\alpha}^i_n)$;
these are the only ones that remain after we set  
$\alpha^\dagger _n=0$ and $\tilde{\alpha}^\dagger _n=0$ (light cone gauge) 
and use the constraints  to determine 
 $\alpha^-_n, \tilde{\alpha}^-_n$ as  functions of $\alpha^i_n, \tilde{\alpha}^i_n$.
 Then the  relevant oscillator number operators are 
\be
N=\sum_{n=1}^\infty \alpha^i_{-n}\alpha^i_{n}, \qquad \qquad 
\bar{N}=\sum_{n=1}^\infty \tilde\alpha^i_{-n}\tilde\alpha^i_{n}
\ee
The mass shell conditions  are
\be
(L_0 -1)|\psi> = 0, \qquad (\bar{L}_0 -1)|\psi> = 0\ .
\ee
The physical states  are 
\be
|\psi> = \xi_{i_1\ldots i_n, j_1\ldots j_k}(p)\ \alpha^{i_1}_{-m_1}\ldots \alpha^{i_n}_{-m_n}
\tilde\alpha^{j_1}_{-s_1}\ldots\tilde\alpha^{j_k}_{-s_k} |0,p> \ , 
\ee
For them 
\be
\alpha' M^2 =2(\ell +\bar{\ell})-4, \qquad \ \ \ \ell -\bar{\ell}=0 \ ,
\ee
\be
\ell= i_1 m_1 +\ldots + i_n m_n, \qquad\ \ \  \bar \ell= j_1 s_1+ \ldots j_k s_k
\ee
i.e. 
$
\alpha'M^2 = 4(\ell-1).
$
We thus  find:

\vspace{0.3cm}
$\ell=0$: $\qquad \alpha'M^2 =-4$ --- \  scalar tachyon

\vspace{0.3cm}
$\ell=1$: $\qquad \alpha'M^2 =0$ --- \ \ massless state

 This   state  $\xi_{ij}(p) \alpha^i_{-1}\tilde\alpha^j_{-1}|0,p>$ 
 can be split into   irreps of $SO(D-2)$:
\be
\xi_{ij} = \bar\xi_{(ij)} + \delta_{ij}\phi + \xi_{[ij]}
\ee
The  symmetric traceless tensor  $\bar\xi_{(ij)}$ 
represents  spin-2 graviton  which
 lies on the leading Regge trajectory. 
 The second term $\phi$ is a scalar dilaton. 
 The antisymmetric tensor corresponding to  $\xi_{[ij]}$ is called the Kalb-Ramond field.

The graviton  has $\frac{(D-2)(D-2+1)}{2}-1=\frac{D(D-3)}{2}$ components, and 
the Kalb-Ramond field  has $\frac{(D-2)(D-3)}{2}$ components
(in $D=4$ the graviton has two degrees of freedom and the Kalb-Ramond field has one, 
i.e. is equivalent to a scalar).

Higher massive  levels are described by $SO(D-2)$ tensors combined in irreps of $SO(D-1)$.

The superstring spectrum has similar structure with the ground state being massless
(representing $D=10$ supergravity states)
and  with fermions as well as bosons as  physical states.

\

Many more details and various extensions can be found in the
books listed below. 

\

{\bf Acknowledgements }
\noindent
These lectures were delivered  17-21 August   2006  at the  School 
«``Physics of Fundamental Interactions'' 
in Protvino   supported by the Dynasty Foundation.

\

\

{\bf Some books on string theory}

\end{document}